\documentclass[sigconf,review=false]{acmart}
\usepackage{booktabs}
\usepackage{multirow}
\usepackage{subfigure}

\usepackage{balance}
\usepackage{flushend}
\usepackage{xspace}
\usepackage{url}
\usepackage{array}
\usepackage{verbatim}
\usepackage{soul, color}
\usepackage{enumitem}
\usepackage{multirow}
\usepackage{amsmath}
\usepackage[ruled,linesnumbered]{algorithm2e}

\theoremstyle{definition}
\newcommand{\eat}[1]{}
\newcommand{\ie}{\emph{i.e.,}\xspace}
\newcommand{\eg}{\emph{e.g.,}\xspace}

\newcommand{\baby}{MBHT\xspace}
\newcommand{\paratitle}[1]{\noindent\textbf{#1}}

\newcommand{\bm}[1]{\boldsymbol{#1}}
\newcommand{\trans}{{\mkern-1.5mu\mathsf{T}}}
\def\model{MBHT}

\keywords{Sequential Recommendation, Graph Neural Networks, Hypergraph Learning, Multi-Behavior Recommendation}

\copyrightyear{2022}
\acmYear{2022}
\setcopyright{acmcopyright}\acmConference[KDD'22]{Proceedings of the 28th ACM SIGKDD Conference on Knowledge Discovery and Data Mining}{August 14--18, 2022}{Washington, DC, USA}
\acmBooktitle{Proceedings of the 28th ACM SIGKDD Conference on Knowledge Discovery and Data Mining (KDD'22), August 14--18, 2022, Washington, DC, USA}
\acmPrice{15.00}
\acmDOI{10.1145/3534678.3539342}
\acmISBN{978-1-4503-9385-0/22/08}


\begin{document}




\title{Multi-Behavior Hypergraph-Enhanced Transformer \\ for Sequential Recommendation}









\author{Yuhao Yang}
\affiliation{
  \institution{University of Hong Kong}
  \city{Hong Kong}
  \country{China}
}
\email{yuhao-yang@outlook.com}

\author{Chao Huang}
\authornote{Chao Huang is the corresponding author.}
\affiliation{
  \institution{University of Hong Kong}
  \city{Hong Kong}
  \country{China}
}
\email{chaohuang75@gmail.com}

\author{Lianghao Xia}
\affiliation{
  \institution{University of Hong Kong}
  \city{Hong Kong}
  \country{China}
}
\email{aka\_xia@foxmail.com}

\author{Yuxuan Liang}
\affiliation{
    \institution{National University of Singapore}
    \city{Singapore}
    \country{Singapore}
}
\email{yuxliang@outlook.com}

\author{Yanwei Yu}
\affiliation{%
  \institution{Ocean University of China}
  \city{Qingdao}
  \country{China}
}
\email{yuyanwei@ouc.edu.cn}

\author{Chenliang Li}
\affiliation{%
   \institution{Wuhan University}
   \city{Wuhan}
   \country{China}
}
\email{cllee@whu.edu.cn}

\renewcommand{\shortauthors}{Yuhao Yang et al.}

\begin{abstract}
Learning dynamic user preference has become an increasingly important component for many online platforms (\eg video-sharing sites, e-commerce systems) to make sequential recommendations. Previous works have made many efforts to model item-item transitions over user interaction sequences, based on various architectures, \eg recurrent neural networks and self-attention mechanism. Recently emerged graph neural networks also serve as useful backbone models to capture item dependencies in sequential recommendation scenarios. Despite their effectiveness, existing methods have far focused on item sequence representation with singular type of interactions, and thus are limited to capture dynamic heterogeneous relational structures between users and items (\eg page view, add-to-favorite, purchase). To tackle this challenge, we design a \underline{M}ulti-\underline{B}ehavior \underline{H}ypergraph-enhanced \underline{T}ransformer framework (\model) to capture both short-term and long-term cross-type behavior dependencies. Specifically, a multi-scale Transformer is equipped with low-rank self-attention to jointly encode behavior-aware sequential patterns from fine-grained and coarse-grained levels. Additionally, we incorporate the global multi-behavior dependency into the hypergraph neural architecture to capture the hierarchical long-range item correlations in a customized manner. Experimental results demonstrate the superiority of our \model\ over various state-of-the-art recommendation solutions across different settings. Further ablation studies validate the effectiveness of our model design and benefits of the new \model\ framework. Our implementation code is released at: \url{https://github.com/yuh-yang/MBHT-KDD22}.
\end{abstract}

\maketitle

\section{Introduction}
\label{sec:intro}


Recommendation models have emerged as the core components of many online applications~\cite{zhang2019deep}, such as social media platforms~\cite{zhou2019online}, video streaming services~\cite{jiang2020aspect} and online retail systems~\cite{wu2018turning}. Due to the highly practical value of sequential behavior modeling in various online platforms, sequential recommendation has been widely adopted in online platforms, with the aim of forecasting future users' interacted item based on their past behavior sequences~\cite{sun2019bert4rec,chang2021surge}.


Generally, in the sequential recommendation scenario, the systems rely on the item sequences to model time-evolving user preferences. Upon this learning paradigm, many sequential recommender systems have been proposed to encode the item dependencies based on various neural techniques and provided stronger performance, \eg recurrent neural recommendation architecture-GRU4Rec~\cite{hidasi2015session} and convolution-based sequence encoder model-Caser~\cite{tang2018caser}. Inspired by the Transformer framework, self-attention has been utilized to capture the item-item pairwise correlations in SASRec~\cite{kang2018sasrec} and BERT4Rec~\cite{sun2019bert4rec}. Recently, graph neural networks (GNNs) have shown effectiveness in sequential recommender systems. Owing to strength of graph convolutions, GNN-based methods (\eg SR-GNN~\cite{wu2019srgnn}, GCSAN~\cite{xu2019gcsan}) are developed to learn item transitions with message passing over the constructed item graph structures.


\begin{figure}
	\includegraphics[width=0.95\linewidth]{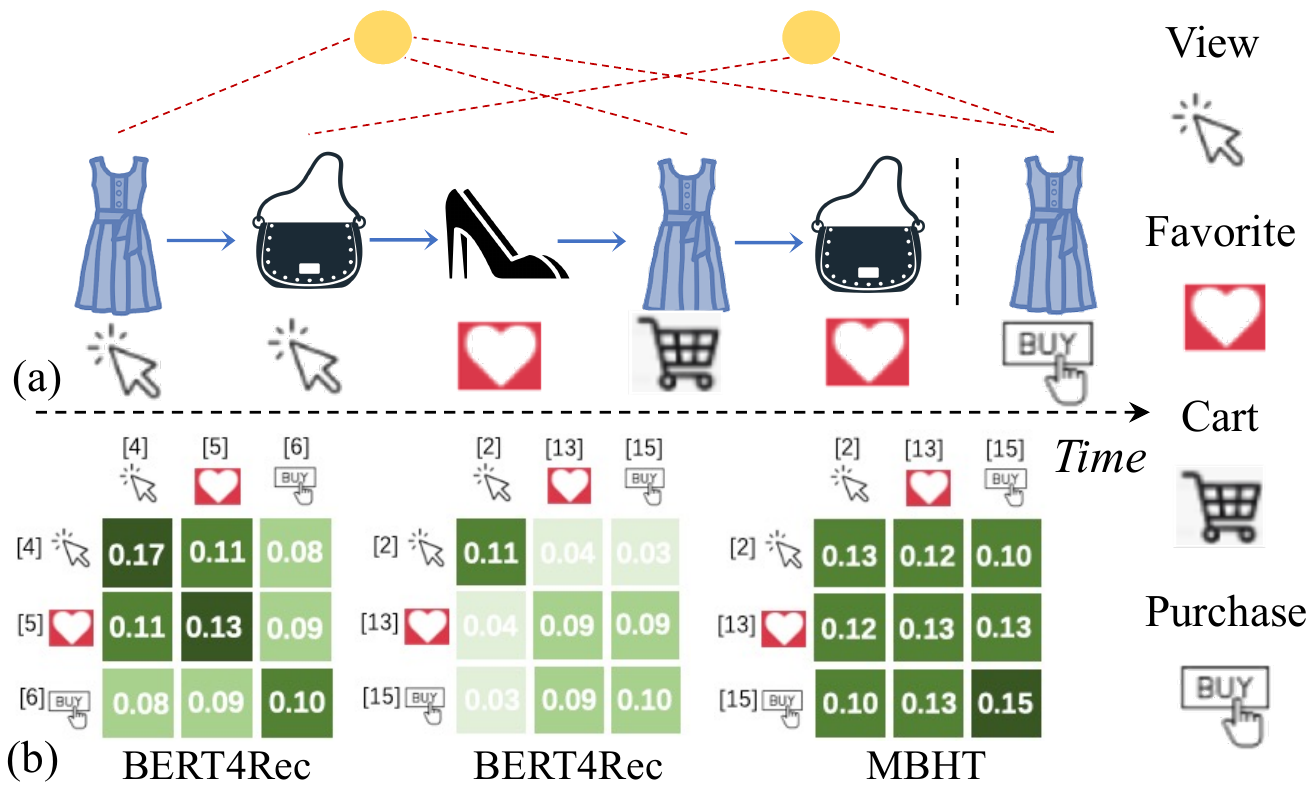}
	\vspace{-0.1in}
	\caption{(a) Illustration example of sequential recommendation with multi-behavior dynamics. (b) Learned behavior-aware dependency weights for short-term item correlations among neighboring $\{[4], [5], [6]\}$ and long-range dependencies among $\{[2], [13], [15]\}$ by BERT4Rec and our \model.}
	\label{fig:intro}
	\vspace{-0.15in}
\end{figure}

Despite their effectiveness, most of existing works have thus far focused on sequential behavior pattern encoding for singular type of interactions (\eg clicks or purchases), without taking multi-typed user-item relationships into consideration. However, in practical online platforms, user behaviors often exhibit both time-dependent and multi-typed, involving different types of user-item interactions, \eg, page view, add-to-favorite, purchase. As illustrated in Figure~\ref{fig:intro}, customers may click and add their interested products to their favorite lists before purchasing them. Additionally, we also visualize the learned dependency weights by the baseline BERT4Rec and our \model\ method to capture behavior-aware short-term item correlations (among neighboring items \{[4], [5], [6]\}) and long-term item-wise dependencies (among \{[2], [13], [15]\}). We can observe that the global multi-behavior dependencies can be better distilled by our \model\ as compared to BERT4Rec. Hence, effectively enhancing the user preference learning with the exploration of heterogeneous user-item interactions in a dynamic environment, is also the key to making accurate sequential recommendations. Nevertheless, this task is not trivial due to the following challenges:
\begin{itemize}[leftmargin=*]
\item \textbf{Dynamic Behavior-aware Item Transitions}. How to explicitly capture the dynamic behavior-aware item transitions with multi-scale temporal dynamics remains a challenge. There exist different periodic behavior patterns (\eg daily, weekly, monthly) for different categories of items (\eg daily necessities, seasonal clothing)~\cite{ren2019lifelong,2019online}. Therefore, it is necessary to explicitly capture multi-scale sequential effects of behavior-aware item transitional patterns from fine-grained to coarse-grained temporal levels. \\\vspace{-0.12in}

\item \textbf{Personalized Global Multi-Behavior Dependencies}. The implicit dependencies across different types of behaviors over time vary from user to user. For example, due to the personalized and diverse user interaction preferences, some people would like to add products to their favorite list if they show interested in the items. Others may prefer to generate their favorite item list with products they are very likely to buy. That is to say, for different users, multi-behaviors have various time-aware dependencies on their interests. Moreover, multi-behavior item-wise dependencies are beyond pairwise relations and may exhibit triadic or event higher-order. Hence, the designed model requires a tailored modeling of diverse users' multi-behavior dependencies with a dynamic multi-order relation learning paradigm.
\end{itemize}

\noindent \textbf{Contribution}. This work proposes a \underline{M}ulti-\underline{B}ehavior \underline{H}ypergraph-enhanced \underline{T}ransformer (\model) to capture dynamic item dependencies with behavior type awareness. Our developed \model\ framework consitens of two key learning paradigms to address the aforementioned challenges correspondingly. (1) \textbf{Behavior-aware Sequential Patterns}. We propose a multi-scale Transformer module to comprehensively encode the multi-grained sequential patterns from fine-grained level to coarse-grained level for behavior-aware item transitions. To improve the efficiency of our sequential pattern encoder, we equip our multi-scale Transformer with the self-attentive projection based on low-rank factorization. To aggregate scale-specific temporal effects, a multi-scale behavior-aware pattern fusion is introduced to integrate multi-grained item transitional signals into a common latent representation space. (2) \textbf{Global Modeling of Diverse Multi-Behavior Dependencies}. We generalize the modeling of global and time-dependent cross-type behavior dependencies with a multi-behavior hypergraph learning paradigm. We construct the item hypergraph structures by unifying the latent item-wise semantic relateness and item-specific multi-behavior correlations. Technically, to capture the global item semantics, we design the item semantic dependence encoder with metric learning. Upon the hypergraph structures, we design the multi-behavior hyperedge-based message passing schema for refining item embeddings, which encourages the long-range dependency learning of different types of user-item relationships. Empirically, \model\ is able to provide better performance than state-of-the-art methods, \eg BERT4Rec~\cite{sun2019bert4rec}, HyperRec~\cite{wang2020next}, SURGE~\cite{chang2021surge}, MB-GMN~\cite{xia2021graph}.



The main contributions are summarized as follows:
\begin{itemize}[leftmargin=*]

\item This work proposes a new framework named \model\ for sequential recommendation, which uncovers the underlying dynamic and multi-behavior user-item interaction patterns.

\item To model multi-grained item transitions with the behavior type awareness, we design a multi-scale Transformer which is empowered with the low-rank and multi-scale self-attention projection, to maintain the evolving relation-aware user interaction patterns.

\item In addition, to capture the diverse and long-range multi-behavior item dependencies, we propose a multi-behavior hypergraph learning paradigm to distill the item-specific multi-behavior correlations with global and customized sequential context injection.

\item We perform extensive experiments on three publicly available datasets, to validate the superiority of our proposed \model\ over various state-of-the-art recommender systems. Model ablation and case studies further show the benefits of our model.

\end{itemize}

\section{Problem Formulation}

In this section, we introduce the primary knowledge and formulates the task of multi-behavior sequential recommendation.\\\vspace{-0.13in}

\noindent \textbf{Behavior-aware Interaction Sequence}.
Suppose we have a sequential recommender system with a set of $I$ users $u_i\in \mathcal{U}$ where $|\mathcal{U}|=I$. For an individual user $u_i$, we define the behavior-aware interaction sequence $S_i=[(v_{i,1}, b_{i,1}),..., (v_{i,j}, b_{i,j}), ...,(v_{i,J}, b_{i,J})]$ with the consideration of item-specific interaction type, where $J$ denotes the length of temporally-ordered item sequence. Here, we define $b_{i,j}$ to represent the behavior type of the interaction between user $u_i$ and $j$-th item $v_j$ in $S_i$, such as page view, add-to-favorite, add-to-cart and purchase in e-commerce platforms.\\\vspace{-0.13in}

\noindent \textbf{Task Formulation}. In our multi-behavior sequential recommender system, different types of interaction behaviors are partitioned into \emph{target behaviors} and \emph{auxiliary behaviors}. Specifically, we regard the interaction with the behavior type we aim to predict as target behaviors. Other types of user behaviors are defined as auxiliary behaviors to provide various behaviour contextual information about users' diverse preference, so as to assist the recommendation task on the target type of user-item interactions. For example, in many online retail platforms, purchase behaviors can be considered as the prediction targets, due to their highly relevance to the Gross Merchandise Volume (GMV) in online retailing to indicate the total sales value for merchandise~\cite{2019online,wu2018turning}. We formally present our studied sequential recommendation problem as follows:\vspace{-0.05in}

\begin{itemize}[leftmargin=*]

\item \textbf{Input}: The behavior-aware interaction sequence $S_i=[(v_{i,1}, b_{i,1})\\,...,(v_{i,j}, b_{i,j}),...,(v_{i,J}, b_{i,J})]$ of each user $u_i\in \mathcal{U}$.\\\vspace{-0.12in}

\item \textbf{Output}: The learning function that estimates the probability of user $u_i$ will interact with the item $v_{J+1}$ with the target behavior type at the future $(J+1)$-th time step.

\end{itemize}

\section{METHODOLOGY}
\begin{figure*}
    \centering
    \includegraphics[width=\textwidth]{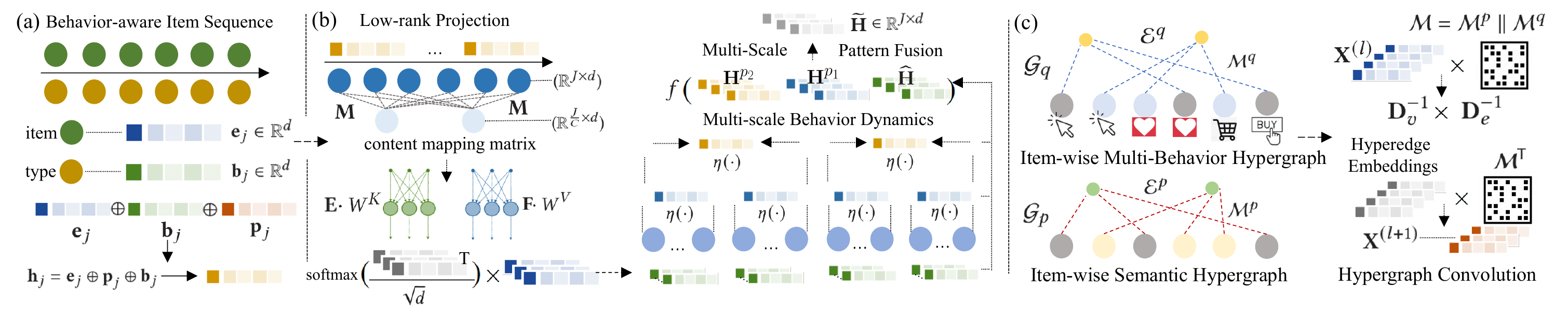}
    \vspace{-0.3in}
    \caption{\model's model flow. (a) We inject the behavior-aware interaction context into item embeddings $\textbf{h}_j = \textbf{e}_j \oplus \textbf{p}_j \oplus \textbf{b}_j$. (b) Multi-scale transformer architecture to capture behavior-aware transitional patterns via low-rank self-attention and  multi-scale sequence aggregation. Scale-specific behavior patterns are fused through the fusion function $\widetilde{\textbf{H}} = f(\widehat{\textbf{H}} \mathbin\Vert {\textbf{H}}^{p_1} \mathbin\Vert {\textbf{H}}^{p_2})$. (c) We capture the global and personalized multi-behavior dependency learning with our hypergraph neural architecture over $\mathcal{G}$.}
    \label{fig:framework}
    \vspace{-0.1in}
\end{figure*}

Figure~\ref{fig:framework} presents the overall architecture of our proposed \model\ model which consists of three key modules: i) Multi-scale modeling of behavior-aware transitional patterns of user preference; ii) Global learning of multi-behavior dependencies of time-aware user interactions; iii) Cross-view aggregation with the encoded representations of sequential behavior-aware transitional patterns and hypergraph-enhanced multi-behavior dependencies.

\subsection{Multi-Scale Modeling of Behavior-aware Sequential Patterns}
In this section, we present the technical details of our \model\ in capturing the behavior-aware user interest with multi-scale dynamics.


\subsubsection{\rm \textbf{Behavior-aware Context Embedding Layer}}
To inject the behavior-aware interaction context into our sequential learning framework, we design the behavior-aware context embedding layer to jointly encode the individual item information and the corresponding interaction behavior contextual signal. Towards this end, given an item $v_j$, we offer its behavior-aware latent representation $\textbf{h}_j \in \mathbb{R}^{d}$ with the following operation:
\begin{align}
	\textbf{h}_j = \textbf{e}_j \oplus \textbf{p}_j \oplus \textbf{b}_j
\end{align}
\noindent where $\textbf{e}_j \in \mathbb{R}^{d}$ represents the initialized item embedding. $\textbf{b}_j \in \mathbb{R}^{d}$ is the behavior type embedding corresponding to the interaction type (\eg page view, add-to-favorite) between user $u_i$ and item $v_j$. Here, $\textbf{p}_j \in \mathbb{R}^{d}$ represents the learnable positional embedding of item $v_j$ which differentiates the temporally-ordered positional information of different interacted items. After this context embedding layer, we can obtain the item representation matrix $\textbf{H} \in \mathbb{R}^{J\times d}$ for the behavior-aware interacted item sequence $S_i$ of user $u_i$.


\subsubsection{\rm \textbf{Multi-Scale Transformer Layer}}
In practical recommendation scenarios, user-item interaction preference may exhibit multi-scale transitional patterns over time. For instance, users often purchase different categories of products (\eg daily necessities, clothes, digital devices) with different periodic trends, such as daily or weekly routines~\cite{jiang2020aspect}. To tackle this challenge, we design a multi-scale sequential preference encoder based on the Transformer architecture to capture the multi-grained behavior dynamics in the behavior-aware interaction sequence $S_i$ of users ($u_i\in \mathcal{U}$).\\\vspace{-0.1in}



\noindent \textbf{Low-Rank Self-Attention Module}. Transformer has shown its effectiveness in modeling relational data across various domains (\eg language~\cite{yao2020multimodal}, vision~\cite{liu2021swin}). In Transformer framework, self-attention serves as the key component to perform the relevance-aware information aggregation among attentive data points (\eg words, pixels, items). However, the high computational cost (\ie quadratic time complexity) of the self-attention mechanism limits the model scalability in practical settings~\cite{kitaev2020reformer}. Motivated by the design of Transformer structure in~\cite{wang2020linformer}, we design a low-rank-based self-attention layer without the quadratic attentive operation, to approximate linear model complexity.

In particular, different from the original scaled dot-product attention for pairwise relation encoding, we generate multiple smaller attention operations to approximate the original attention with low-rank factorization. We first define two trainable projection matrices $\boldsymbol{E} \in \mathbb{R}^{\frac{J}{C} \times J}$ and $\boldsymbol{F} \in \mathbb{R}^{\frac{J}{C} \times J}$ to perform the low-rank embedding transformation. Here, $C$ denotes the low-rank scale and $\frac{J}{C}$ represents the number of low-rank latent representation spaces over the input behavior-aware interaction sequence $S_i$. Formally, we represent our low-rank self-attention as follows:
\begin{align}
\label{eq:lowrank}
		\widehat{\textbf{H}} =\textsf{softmax}(\frac{\textbf{H}  \cdot {W}^Q(\textbf{E} \cdot \textbf{H}  \cdot {W}^K)^\trans}{\sqrt{d}})\cdot \textbf{F} \cdot \textbf{H}  \cdot {W}^V
\end{align}
\noindent where ${W}^Q$, ${W}^K$, ${W}^V$ are learnable transformation matrices for embedding projection. In our low-rank self-attention module, $\boldsymbol{E}$ and $\boldsymbol{F}$ are utilized to project the ($\mathbb{R}^{J\times d}$)-dimensional key and value transformed representations $\textbf{H}  \cdot {W}^K$ and $\textbf{H}  \cdot {W}^V$ into ($\mathbb{R}^{\frac{J}{C}\times d}$)-dimensional latent low-rank embeddings $\widehat{\textbf{H}} \in \mathbb{R}^{\frac{J}{C}\times d}$. In summary, with the low-rank factor decomposition over the original attention operations, we calculate the context mapping matrix $\textbf{M}=\frac{\textbf{H}  \cdot {W}^Q (\textbf{H}  \cdot {W}^K)^T}{\sqrt{d}}$ with the dimension of $\mathbb{R}^{J\times \frac{J}{C}}$ as compared to the original dimension $\mathbb{R}^{J\times J}$ in the vanilla self-attention mechanism. By doing so, the computational cost of our behavior sequence encoder can be significantly reduced from the $O(J\times J)$ to $O(J\times \frac{J}{C})$ given that the low-rank projected dimension $J/C$ is often much smaller than $J$, \ie $J/C \ll\ J$.\\\vspace{-0.12in}


\paratitle{Multi-Scale Behavior Dynamics.}
To endow our \model\ model with the effective learning of multi-scale behaviour transitional patterns, we propose to enhance our low-rank-based transformer with a hierarchical structure, so as to capture granularity-specific behavior dynamics. To be specific, we develop a granularity-aware aggregator to generate  granularity-specific representation $\textbf{g}_p$ which preserves the short-term behavior dynamic. Here, we define $p$ as the length of sub-sequence for a certain granularity. We formally present our granularity-aware emebdding generation with the aggregated representation $\boldsymbol{\Gamma}^p \in \mathbb{R}^{\frac{J}{p} \times d}$ and $\boldsymbol{\gamma} \in \mathbb{R}^d$ as follows:
\begin{align}
\label{eq:attn_pool}
\boldsymbol{\Gamma}^p = \{\boldsymbol{\gamma}_1,...,\boldsymbol{\gamma}_{\frac{J}{p}}\} = [\eta(\textbf{h}_1,...,\textbf{h}_p);...; \eta(\textbf{h}_{J-p+1},...,\textbf{h}_{J})]
\end{align}
\noindent where $\eta(\cdot)$ represents the aggregator to capture the short-term behavior-aware dynamics. Here, we utilize the mean pooling to perform the embedding aggregation. After that, we feed the granularity-aware behavior representations into a self-attention layer for encoding granularity-specific behavior pattern as shown below:
\begin{align}
\label{eq:pool_attn}
		{\textbf{H}}^p =\textsf{softmax}(\frac{\boldsymbol{\Gamma}^p \cdot {W}^Q_p(\boldsymbol{\Gamma}^p \cdot {W}^K_p)^\trans}{\sqrt{d}}) \cdot \boldsymbol{\Gamma}^p \cdot {W}^V_p
\end{align}
\noindent ${\textbf{H}}^p \in \mathbb{R}^{\frac{J}{p} \times d}$ encodes the short-term transitional patterns over different item sub-sequences. In our \model\ framework, we design a hierarchical Transformer network with two different scale settings $p_1$ and $p_2$. Accordingly, our multi-scale Transformer can produce three scale-specific sequential behavior embeddings $\widehat{\textbf{H}} \in \mathbb{R}^{J\times d}$, ${\textbf{H}}^{p_1} \in \mathbb{R}^{\frac{J}{p_1} \times d}$, ${\textbf{H}}^{p_2} \in \mathbb{R}^{\frac{J}{p_2} \times d}$.


\subsubsection{\bf Multi-Scale Behaviour Pattern Fusion}
To integrate the multi-scale dynamic behavior patterns into a common latent representation space, we propose to aggregate the above encoded scale-specific embeddings with a fusion layer presented as follows:
\begin{align}
\widetilde{\textbf{H}} = f(\widehat{\textbf{H}} \mathbin\Vert {\textbf{H}}^{p_1} \mathbin\Vert {\textbf{H}}^{p_2})
\end{align}
\noindent Here $f(\cdot)$ represents the projection function which transforms $\mathbb{R}^{(\frac{J}{C} + \frac{J}{p_1} + \frac{J}{p_2}) \times d}$ dimensional embeddings into $\mathbb{R}^{J\times d}$ dimensional representations corresponding to different items ($v_j\in S_i$) in the behavior-aware interaction sequence $S_i$ of user $u_i$. Here, $\parallel$ denotes the concatenation operation over different embedding vectors. \\\vspace{-0.12in}


\paratitle{Multi-Head-Enhanced Representation Spaces.}
In this part, we propose to endow our behavior-aware item sequence encoder with the capability of jointly attending multi-dimensional interaction semantics. In particular, our multi-head sequential pattern encoder projects the $\textbf{H}$ into $N$ latent representation spaces and performs head-specific attentive operations in parallel.
\begin{align}
	\widetilde{\textbf{H}} &= (\text{head}_1 \mathbin\Vert \text{head}_2 \mathbin\Vert  \cdots \mathbin\Vert \text{head}_N)\boldsymbol{W}^D \\\nonumber
	\text{head}_n &= f(\widehat{\textbf{H}}_n \mathbin\Vert {\textbf{H}}_n^{p_1} \mathbin\Vert {\textbf{H}}_n^{p_2})
\end{align}
\noindent where $\widehat{\textbf{H}}_n, {\textbf{H}}_n^{p_1}$ and ${\textbf{H}}_n^{p_2}$ are computed with head-specific projection matrices $W_n^Q, W_n^K, W_n^V \in \mathbb{R}^{d\times d/N}$, and $\boldsymbol{W}^D \in \mathbb{R}^{d\times d}$ is the output transformation matrix. The multiple attention heads allows our multi-scale Transformer architecture to encode multi-dimensional dependencies among items in $S_i$.\\\vspace{-0.12in}


\paratitle{Non-linearity Injection with Feed-forward Module.}
In our multi-scale Transformer, we use the point-wise feed-forward network to inject non-linearities into the new generated representations. The non-linear transformation layer is formally represented:
\begin{align}
	\textsf{PFFN}(\widetilde{\textbf{H}}^{(l)}) &= [\textsf{FFN}(\tilde{\textbf{h}}_1^{(l)})^\trans, \cdots, \textsf{FFN}(\tilde{\textbf{h}}_t^{(l)})^\trans] \\\nonumber
	\textsf{FFN}(\textbf{x}) &= \textsf{GELU}(\textbf{x}\textbf{W}_1^{(l)} + \textbf{b}_1^{(l)})\textbf{W}_2^{(l)}+\textbf{b}_2^{(l)},
\end{align}
\noindent In our feed-forward module, we adopt two layers of non-linear transformation with the integration of intermediate non-linear activation $\textsf{GELU}(\cdot)$. In addition, $\textbf{W}_1^{(l)} \in \mathbb{R}^{d \times d_h}, \textbf{W}_2^{(l)} \in \mathbb{R}^{d_h \times d}, \textbf{b}_1^{(l)} \in \mathbb{R}^{d}, \textbf{b}_2^{(l)} \in \mathbb{R}^{d}$ are learnable parameters of projection matrices and bias terms. Here, $l$ denotes the $l$-th multi-scale Transformer layer.


\subsection{Customized Hypergraph Learning of Global Multi-Behavior Dependencies}
\label{sec:hg}
In our \model\ framework, we aim to incorporate long-range multi-behavior dependency into the learning paradigm of evolving user interests. However, it is non-trivial to effectively capture the personalized long-range multi-behavior dependencies. To achieve our goal, we propose to tackle two key challenges in our learning paradigm: 

\begin{itemize}[leftmargin=*]
\item i) \textbf{Multi-Order Behavior-wise Dependency}. The item-wise multi-behavior dependencies are no longer dyadic with the consideration of comprehensive relationships among different types of user behaviors. For example, when deciding to recommend  a specific item to users for their potential purchase preference, it would be useful to explore past multi-behavior interactions (\eg page view, add-to-favorite) between users and this item. Customers are more likely to add their interested products into their favorite item list before making final purchases.\\\vspace{-0.12in}

\item ii) \textbf{Personalized Multi-Behavior Interaction Patterns}. Multi-behavior patterns may vary by users with different correlations across multi-typed user-item interactions. In real-life e-commerce systems, some users like to add many items to their favorite list or cart, if they are interested in, but only a few of them will be purchased later. In contrast, another group of users only tag their interested products as favorite only if they show strong willingness to buy them. Hence, such complex and personalized multi-behavior patterns require our model to preserve the diverse cross-type behaviour dependencies.

\end{itemize}

To address the above challenges, we build our global multi-behavior dependency encoder upon the hypergraph neural architecture. Inspired by the flexibility of hypergraphs in connecting multiple nodes through a single edge~\cite{feng2019hypergraph,xia2022hypergraph}, we leverage the hyperedge structure to capture the tetradic or higher-order multi-behavior dependencies over time. Additionally, given the behavior-aware interaction sequence of different users, we construct different hypergraph structures over the sequence $S_i$ ($u_i\in \mathcal{U}$), with the aim of encoding the multi-behavior dependency in a customized manner.

\subsubsection{\bf Item-wise Hypergraph Construction}
In our hypergraph framework, we generate two types of item-wise hyperedge connections corresponding to i) long-range semantic correlations among items; ii) item-specific multi-behavior dependencies across time.\\\vspace{-0.12in}

\label{sec:hg_construct}
\noindent \textbf{Item Semantic Dependency Encoding with Metric Learning}.
To encode the time-evolving item semantics and the underlying long-range item dependencies based on the same user interest (\eg food, outdoor activities), we introduce an item semantic encoder based on a metric learning framework. Specifically, we design the learnable metric $\hat{\beta}^n_{j,j'}$ between items with a multi-channel weight function $\tau({\cdot})$ presented as follows:
\begin{align}
	\beta_{j,j'} & = \frac{1}{N}\sum_{n=1}^{N} \hat{\beta}^n_{j,j'}; \textbf{v}_j = \textbf{e}_j \oplus \textbf{b}_j \\
	\hat{\beta}^n_{j,j'} & = \tau(\boldsymbol{w}^n \odot \textbf{v}_j, \boldsymbol{w}^n \odot \textbf{v}_{j'})
\end{align}
\noindent where $\hat{\beta}^n_{j,j'}$ represents the learnable channel-specific dependency weight between item $v_j$ and $v_{j'}$. We define the weight function $\tau({\cdot})$ as the cosine similarity estimation based on the trainable $\boldsymbol{w}^n$ of $n$-th representation channel. ($\boldsymbol{w}^n \odot \textbf{v}_{j}$) represents the embedding projection operation. In our item-wise semantic dependency, we perform metric learning under $N$ representation channels (indexed by $n$). The mean pooling operation is applied to all learned channel-specific item semantic dependency scores (\eg $\hat{\beta}^n_{j,j'}$), to obtain the final relevance $\beta_{j,j'}$ between item $v_j$ and $v_{j'}$.\\\vspace{-0.12in}

\noindent \textbf{Item-wise Semantic Dependency Hypergraph}. With the encoded semantic dependencies among different items, we generate the item-wise semantic hypergraph by simultaneously connecting multiple highly dependent items with hyperedges. In particular, we construct a set of hyperedges $\mathcal{E}^p$, where $|\mathcal{E}^p|$ corresponds to the number of unique items in sequence $S_i$. In the hypergraph $\mathcal{G}_p$ of item-wise semantic dependencies, each unique item will be assigned with a hyperedge in $\epsilon \in \mathcal{E}^p$ to connect top-$k$ semantic dependent items according to the learned item-item semantic dependency score $\beta_{j,j'}$ (encoded from the metric learning component). $A_j$ represents the set of top-$k$ semantic correlated items of a specific item $v_j$. We define the connection matrix between items and hyperedges as $\mathcal{M}^p \in \mathbb{R}^{J\times |\mathcal{E}^p|}$ in which each entry $m^{p}({v_j,\epsilon_{j'}})$ is:
\begin{align}
\label{eq:sim_edge}
	m^{p}({v_j,\epsilon_{j'}}) &= \begin{cases}
		\beta_{j,j'} & v_{j'} \in A_j; \\
		0 & otherwise;
	\end{cases}
\end{align}
\noindent where $\epsilon_{j'}$ denotes the hyperedge which is assigned to item $v_{j'}$.\\\vspace{-0.12in}


\noindent \textbf{Item-wise Multi-Behavior Dependency Hypergraph}.
To capture the personalized item-wise multi-behavior dependency in a time-aware environment, we further generate a hypergraph structure $\mathcal{G}_q$ based on the observed multi-typed interactions (\eg page view, add-to-cart) between user $u_i$ and a specific item $v_j$ at different timestamps. Here, we define $\mathcal{E}^q$ to represent the set of items which have multi-typed interactions with user $u_i$. In hypergraph $\mathcal{G}_q$, the number of hyperedges is equal to $|\mathcal{E}^q|$. Given that users have diverse multi-behaviour patterns with their interacted items, the constructed multi-behavior dependency hypergraphs vary by users. To be specific, we generate item-hyperedge connection matrix $\mathcal{M}^q \in \mathbb{R}^{J\times |\mathcal{E}^q|}$ ($m^q \in \mathcal{M}^q$) for hypergraph $\mathcal{G}_q$ as follow:
\begin{align}
\label{eq:mb_edge}
	m^{q}(v_j^b,\epsilon_{j}^{b'}) &= \begin{cases}
		1 & v_{j}^b \in \mathcal{E}^q_j; \\
		0 & otherwise;
	\end{cases}
\end{align}
\noindent $v_j^b$ represents that item $v_j$ is interacted with user $u_i$ under the $b$-th behavior type. $\mathcal{E}^q_j$ denotes the set of multi-typed $u_i$-$v_j$ interactions.



We further integrate our constructed hypergraph structures $\mathcal{G}_p$ and $\mathcal{G}_q$ by concatenating connection matrices $\mathcal{M}^p$ and $\mathcal{M}^q$ along with the column side. As such, the integrated hypergraph $\mathcal{G}$ is constructed with the concatenated connection matrix $\mathcal{M} \in \mathbb{R}^{J \times (|\mathcal{E}^p|+|\mathcal{E}^q|)}$, \ie $\mathcal{M} = \mathcal{M}^p \mathbin\Vert \mathcal{M}^q$. Different behavior-aware interaction sequences result in different hypergraph structures for different users, which allow our \model\ model to encode the personalized multi-behavior dependent patterns in a customized way.

\subsubsection{\bf Hypergraph Convolution Module}
\label{sec:conv_and_ro}
In this module, we introduce our hypergraph message passing paradigm with the convolutional layer, to capture the global multi-behavior dependencies over time. The hypergraph convolutional layer generally involves two-stage information passing \cite{feng2019hypergraph}, \ie node-hyperedge and hyperedge-node embedding propagation along with the hypergraph connection matrix $\mathcal{M}$ for refining item representations. Particularly, we design our hypergraph convolutional layer as:
\begin{align}
\label{eq:HGCN}
	\textbf{X}^{(l+1)} = \textbf{D}_v^{-1} \cdot \boldsymbol{\mathcal{M}} \cdot \textbf{D}_e^{-1} \cdot \boldsymbol{\mathcal{M}}^\trans \cdot \textbf{X}^{(l)}
\end{align}
\noindent where $\textbf{X}^{(l)}$ represents the item embeddings encoded from the $l$-th layer of hypergraph convolution. Furthermore, $\textbf{D}_v$ and $\textbf{D}_e$ are diagonal matrices for normalization based on vertex and edge degrees, respectively. Note that the two-stage message passing by $\boldsymbol{\mathcal{M}} \cdot \boldsymbol{\mathcal{M}}^\trans$ takes $O((|\mathcal{E}^p|+ |\mathcal{E}^q|)\times J^2)$ calculations, which is quite time-consuming. Inspired by the design in \cite{yu2021self}, we calculate a matrix $\boldsymbol{\mathcal{M}}^\prime$ by leveraging pre-calculated $\beta_{j,j^\prime}$ to obtain a close approximation representation of $\boldsymbol{\mathcal{M}} \cdot \boldsymbol{\mathcal{M}}^\trans$ and thus boost the inference. The detailed process can be found in the supplementary material. We also remove the non-linear projection following \cite{he2020lightgcn} to simplify the message passing process. Each item embedding $\textbf{x}^{(0)}$ in $\textbf{X}^{(0)}$ is initialized with the behavior-aware self-gating operation as: $\textbf{x}^{(0)} = (\textbf{v}_j \oplus \textbf{b}_j) \odot \text{sigmoid} ( (\textbf{v}_j \oplus \textbf{b}_j) \cdot \textbf{w} + \textbf{r})$.

\subsection{Cross-View Aggregation}
In the forecasting layer of \model\ framework, we propose to fuse the learned item representations from different views: 1) multi-scale behavior-aware sequential patterns with Transformer architecture; 2) personalized global multi-behavior dependencies with Hypergraph framework. To enable this cross-view aggregation in an adaptive way, we develop an attention layer to learn explicit importance for view-specific item embeddings. Formally, the aggregation procedure is presented as follows:
\begin{align}
	\label{eq:fuse}
	\alpha_i = \textsf{Attn}(\bm{e}_i) = \frac{\exp(\bm{a}^\trans \cdot \bm{W}_a\bm{e}_i)}{\sum_i \exp(\bm{a}^\trans \cdot \bm{W}_a\bm{e}_i)} \\\nonumber
	\bm{e}_i \in \{ \tilde{\textbf{h}}_i, \tilde{\textbf{x}}_i\};~~\textbf{g}_i = \alpha_1 \cdot \tilde{\textbf{h}}_i \oplus \alpha_2 \cdot \tilde{\textbf{x}}_i
\end{align}
\noindent where $\tilde{\textbf{h}}_i$ and $\tilde{\textbf{x}}_i$ are embeddings from the two views separately for item at the $i$-th position, and $\bm{a} \in \mathbb{R}^{d}$, $\bm{W}_a \in \mathbb{R}^{d\times d}$ are trainable parameters. Here, to eliminate the over-smooth effect of graph convolution, $\tilde{\textbf{x}}_i$ is the average of $\textbf{x}^{(l)}$ across all convolutional layers. Finally, the probability of $i$-th item in the sequence being item $v_j$ is estimated as: $\hat{y}_{i,j}=\textbf{g}_i^\trans \textbf{v}_j$, where $\textbf{v}_j$ represents item $v_j$'s embedding.


\subsection{Model Learning And Analysis}
To fit our multi-behavior sequential recommendation scenario, we utilize the Cloze task~\cite{kang2021entangled,sun2019bert4rec} as our training objective to model the bidirectional information of item sequence. We describe our Cloze task settings as follows: Considering the multi-behavior sequential recommender systems, we mask all items in the sequence with the target behavior type (\eg purchase). To avoid the label leak issue, we replace the masked items as well as the corresponding behavior type embeddings with the special token \textsf{[mask]}, and leave out masked items in the hypergraph construction in Section \ref{sec:hg_construct}. Instead, for masked items, we generate its hypergraph embedding $\textbf{x}_i$ in Equation \ref{eq:fuse} by adopting sliding-window average pooling function over hypergraph embeddings of the contextual neighbors surrounding the mask position $(m-q_1, m+q_2)$. The details are presented in Algorithm \ref{algorithm} in the supplementary material. Hence, our model makes prediction on masked items based on the encoded  surrounding context embeddings in the behavior interaction sequence. Given the probability estimation function $\hat{y}_{i,j}=\textbf{g}_i^\trans \textbf{v}_j$, we define our optimized objective loss with Cross-Entropy as below:
\begin{align}
	\mathcal{L} = \frac{1}{|T|} \sum_{t \in T, m \in M} -\log(\frac{\exp \hat{y}_{m,t}}{\sum_{j \in V} \exp \hat{y}_{m,j}})
\end{align}
\noindent where $T$ is the set of ground-truth ids for masked items in each batch, $M$ is the set of masked positions corresponding to $T$, and $V$ is the item set. The time complexity is analyzed in supplementary material.

\vspace{-0.05in}
\section{Experiments}
\begin{table}[t]
\centering
\caption{Statistical information of experimented datasets.}
\vspace{-0.15in}
\label{tab:datasets}
\resizebox{\linewidth}{!}{
\begin{tabular}{c|c|c|c}
\toprule
Stats.          & Taobao & Retailrocket & IJCAI  \\
\hline
\# Users        & 147, 892 & 11, 649 & 200, 000        \\
\# Items        & 99, 038 & 36, 223 & 808, 354        \\
\# Interactions & 7, 092, 362 & 87, 822 & 13, 072, 940 \\
\# Average Length & 48.23 & 14.55 & 78.58 \\
\# Density        & $5\times 10^{-6}$ & $1\times 10^{-6}$ & $7\times 10^{-7}$ \\
\# Behavior Types & [buy, cart, fav, pv] & [buy, cart, pv] & [buy, cart, fav, pv] \\
\bottomrule
\end{tabular}
\vspace{-0.25in}
}
\end{table}


This section aims to answer the following research questions:
\begin{itemize}[leftmargin=*]
\item \textbf{RQ1}: How does our \model\ perform as compared to various state-of-the-art recommendation methods with different settings?
\item \textbf{RQ2}: How effective are the key modules (\eg multi-scale attention encoder, multi-behavior hypergraph learning) in \model? 
\item \textbf{RQ3}: How does \model\ perform to alleviate the data scarcity issue of item sequences when competing with baselines?
\item \textbf{RQ4}: How do different hyperparameters affect the model performance? (Evaluation results are presented in Appendix \ref{sec:hp}).
\end{itemize}


\begin{table*}[t]
	\centering
	\caption{The performance of our method and the best performed baseline are presented with bold and underlined, respectively. Superscript $\ast$ indicates the significant improvement between our \model\ and the best performed baseline with $p$ value $<0.01$.}
	\vspace{-0.1in}
	\label{tab:results}
	\resizebox{\linewidth}{!}{
	\begin{tabular}{c|ccccc|ccccc|ccccc}
		\hline
		\multirow{2}{*}{Model} & \multicolumn{5}{c|}{Taobao} & \multicolumn{5}{c|}{Retailrocket} & \multicolumn{5}{c}{IJCAI} \\
		~ & HR@5 & NDCG@5 & HR@10 & NDCG@10 & MRR & HR@5 & NDCG@5 & HR@10 & NDCG@10 & MRR & HR@5 & NDCG@5 & HR@10 & NDCG@10 & MRR \\
		\hline
		\multicolumn{7}{l}{\textit{General Sequential Recommendation Methods}}\\
		\hline
		Caser & 0.082 & 0.058 & 0.123 & 0.071 & 0.070 & 0.632 & 0.539 & 0.754 & 0.578 & 0.535 & 0.134 & 0.092 & 0.167 & 0.104 & 0.109 \\ 
		HPMN & 0.162 & 0.130 & 0.219 & 0.141 & 0.139 & 0.664 & 0.633 & 0.711 & 0.587 & 0.602 & 0.144 & 0.085 & 0.197 & 0.124 & 0.123\\
		GRU4Rec & 0.147 & 0.105 & 0.209 & 0.125 & 0.118 & 0.640 & 0.575 & 0.708 & 0.597 & 0.572 & 0.141 & 0.100 & 0.200 & 0.119 & 0.113 \\
		SASRec & 0.150 & 0.110 & 0.206 & 0.128 & 0.123 & 0.669 & 0.644 & 0.689 & 0.650 & 0.645 & 0.146 & 0.110 & 0.191 & 0.124 & 0.122 \\
		BERT4Rec & 0.198 & 0.153 & 0.254 & 0.171 & 0.163 & 0.808 & 0.670 & 0.881 & 0.694 & 0.639 & \underline{0.297} & \underline{0.220} & \underline{0.402} & \underline{0.253} & \underline{0.227} \\
		\hline
		\multicolumn{7}{l}{\textit{Graph-based Sequential Recommender Systems}}\\
		\hline
		SR-GNN & 0.102 & 0.071 & 0.153 & 0.087 & 0.086 & 0.848 & 0.780 & 0.891 & 0.793 & 0.767 & 0.072 & 0.048 & 0.118 & 0.062 & 0.064 \\
		GCSAN & \underline{0.217} & 0.160 & 0.305 & \underline{0.188} & 0.173 & 0.872 & 0.846 & 0.890 & 0.851 & 0.842 & 0.119 & 0.086 & 0.175 & 0.104 & 0.101  \\
		HyperRec & 0.145 & 0.130 & 0.224 & 0.133 & 0.129 & 0.860 & 0.705 & 0.833 & 0.820 & 0.816 & 0.140 & 0.109 & 0.236 & 0.144 & 0.132\\
		SURGE & 0.122 & 0.078 & 0.193 & 0.100 & 0.093 & \underline{0.878} & \underline{0.879} & \underline{0.906} & \underline{0.887} & \underline{0.870} & 0.226 & 0.159  & 0.322 & 0.190 & 0.171 \\
		\hline
		\multicolumn{7}{l}{\textit{Multi-Behavior Recommendation Models}}\\
		\hline
		BERT4Rec-MB & 0.211 & \underline{0.169} & 0.263 & 0.186 & \underline{0.178} & 0.875 & 0.858 & 0.889 & 0.863 & 0.857 & 0.257 & 0.189 & 0.342 & 0.216 & 0.197 \\
		MB-GCN & 0.185 & 0.103 & 0.309 & 0.143 & 0.149 & 0.844 & 0.735 & 0.878 & 0.752 & 0.739 & 0.218 & 0.145 & 0.335 & 0.182 & 0.177\\
		NMTR & 0.125 & 0.082 & 0.174 & 0.097 & 0.103 & 0.827 & 0.697 & 0.858 & 0.724 & 0.741 & 0.109 & 0.076 & 0.184 & 0.099 & 0.106\\
		MB-GMN & 0.196 & 0.115 & \underline{0.319} & 0.154 & 0.151 & 0.853 & 0.762 & 0.901 & 0.830 & 0.822 & 0.235 & 0.161 & 0.337 & 0.193 & 0.176\\
		\hline
		\textbf{\baby} & \textbf{0.323}$^\ast$ & \textbf{0.257}$^\ast$ & \textbf{0.405}$^\ast$ & \textbf{0.283}$^\ast$ & \textbf{0.262}$^\ast$ & \textbf{0.931}$^\ast$ & \textbf{0.933}$^\ast$ & \textbf{0.956}$^\ast$ & \textbf{0.950}$^\ast$ & \textbf{0.929}$^\ast$ & \textbf{0.346}$^\ast$ & \textbf{0.268}$^\ast$ & \textbf{0.437}$^\ast$ & \textbf{0.297}$^\ast$ & \textbf{0.272}$^\ast$ \\
		\# Improve & 48.84\% & 52.07\% & 26.95\% & 50.53\% & 47.19\% & 6.04\% & 6.14\% & 5.52\% & 7.10\% & 6.78\% & 16.50\% & 21.82\% & 8.71\% & 17.39\% & 19.82\% \\
		\hline
	\end{tabular}
	}
	\vspace{-0.1in}
\end{table*}

\vspace{-0.05in}
\subsection{Experimental Settings}
\subsubsection{\rm \textbf{Datasets.}}
We utilize three recommendation datasets collected from real-world scenarios. i) \paratitle{Taobao}. This dataset is collected from Taobao which is one of the largest e-commerce platforms in China. Four types of user-item interactions are included in this dataset, \ie target behaviors-\textit{purchase}; auxiliary behaviors-\textit{add-to-favorites, add-to-cart, page view}. ii) \paratitle{Retailrocket}. This dataset is generated from an online shopping site-Retailrocket over 4 months, to record three types of user behaviors, \ie  target behaviors-\textit{purchase}; auxiliary behaviors-\textit{page view \& add-to-cart}. iii) \paratitle{IJCAI}. This dataset is released by IJCAI Contest 2015 for the task of repeat buyers prediction. It shares the same types of interaction behaviors with the Taobao dataset. The detailed statistical information of these experimented datasets are summarized in Table~\ref{tab:datasets}. Note that different datasets vary by average sequence length and user-item interaction density, which provides diverse evaluation settings.






\vspace{-0.05in}
\subsubsection{\rm \textbf{Evaluation Protocols.}}
In our experiments, closely following the settings in~\cite{sun2019bert4rec,tang2018caser}, we adopt the \textit{leave-one-out} strategy for performance evaluation. For each user, we regard the temporally-ordered last purchase as the test samples, and the previous ones as validation samples. Additionally, we pair each positive sample with 100 negative instances based on item popularity~\cite{sun2019bert4rec}. We utilize three evaluation metrics: \textit{Hit Ratio (HR@N), Normalized Discounted Cumulative Gain (NDCG@N)} and \textit{Mean Reciprocal Rank (MRR)}~\cite{wang2020next,yang2022knowledge,wu2019srgnn}. Note that larger HR, NDCG and MRR scores indicate better recommendation performance.

\vspace{-0.05in}
\subsubsection{\rm \textbf{Baselines.}}
We compare our \baby with a variety of recommendation baselines to validate the performance superiority.
\paratitle{General Sequential Recommendation Methods.}
\vspace{-0.05in}
\begin{itemize}[leftmargin=*]
\item \textbf{GRU4Rec}~\cite{hidasi2015session}. It utilizes the gated recurrent unit as sequence encoder to learn dynamic preference with a ranking-based loss.

\item \textbf{SASRec}~\cite{kang2018sasrec}. The self-attention mechanism is leveraged in this method to encode the item-wise sequential correlations.

\item \textbf{Caser}~\cite{tang2018caser}. This method integrates the convolutional neural layers from both vertical and horizontal views to encode time-evolving user preference of item sequence.

\item \textbf{HPMN}~\cite{ren2019lifelong}. It employs a hierarchically structured periodic memory network to model multi-scale transitional information of user sequential behaviors. The incremental updating mechanism is introduced to retain behaviour patterns over time.

\item \textbf{BERT4Rec}~\cite{sun2019bert4rec}. It uses a bidirectional encoder for modeling sequential information with Transformer. The model is optimized with the Cloze objective, and has produced state-of-the-art performance among sequence learning-based baselines.

\end{itemize}

\paratitle{Graph-based Sequential Recommender Systems.}
\vspace{-0.05in}
\begin{itemize}[leftmargin=*]
\item \textbf{SR-GNN}~\cite{wu2019srgnn}. It generates graph structures based on item-item transitional relations in sequences, and conducts graph-based message passing to capture local and global user interests.
\item \textbf{GCSAN}~\cite{xu2019gcsan}. It empowers the self-attention mechanism with with a front-mounted GNN structure. The attentive aggregation is performed over the encoded graph embeddings.
\item \textbf{HyperRec}~\cite{wang2020next}. It designs sequential hypergraphs to capture evolving users' interests and regards users as hyperedges to connect interacted items, so as to model dynamic user preferences. 
\item \textbf{SURGE}~\cite{chang2021surge}. It adopts metric learning to build personalized graphs and uses hierarchical attention to capture multi-dimensional user interests in the graph. 
\end{itemize}
\paratitle{Multi-Behavior Recommendation Models.}
\vspace{-0.05in}
\begin{itemize}[leftmargin=*]
	\item \textbf{BERT4Rec-MB} \cite{sun2019bert4rec}. We enhance the BERT4Rec method to handle the dynamic multi-behavior context by injecting behavior type representations into the input embeddings for self-attention.
	\item \textbf{MB-GCN}~\cite{jin2020mbgcn}. This model is built upon the graph convolutional layer to refine user/item embeddings through the behavior-aware message passing on the user-item interaction graph.
\item \textbf{NMTR}~\cite{gao2019nmtr}. It defines the behavior-wise cascading relationships to model the dependency among different types of behaviors under a multi-task learning paradigm.
	\item \textbf{MB-GMN}~\cite{xia2021graph}. It employs a graph meta network to capture personalized multi-behavior signals and model the diverse multi-behavior dependencies. It generates state-of-the-art performance among different multi-behavior recommendation methods.
\end{itemize}




\subsubsection{\rm \textbf{Hyperparameter Settings.}}
In \model\ model, we search the number of hypergraph propagation layers from \{1,2,3,4\}. The number of multi-head channels is set as 2 for both self-attention and metric learning components. The value of $k$ for constructing item-wise semantic dependency hypergraph is tuned from $[4,6,8,10,12,14]$. Considering that datasets vary by average sequence length, the multi-scale setting parameters $(C, p_1, p_2)$ are searched amongst the value range of ([20,4,20],[20,8,40],[40,4,20],[40,8,40]).


\subsection{Performance Evaluation (RQ1)}
We report the detailed performance comparison on different datasets in Table~\ref{tab:results} and summarize the observations as followed:

\begin{itemize}[leftmargin=*]
\item The proposed \model\ consistently outperforms all types of baselines by a significant margin on different datasets. The performance improvements can be attributed from: i) Through the multi-scale behavior-aware Transformer, \model\ is able to capture the behavior-aware item transitional patterns from fine-grained to coarse-grained time granularities. ii) With the hypergraph neural network for multi-behavior dependency learning, we endow \model\ with the capability of capturing long-range item correlations across behavior types over time. \\\vspace{-0.12in}


\item By jointly analyzing the results across different datasets, we can observe that our \model\ is robust to different recommendation scenarios with various data characteristics, such as average sequence length and user-item interaction density, reflecting various user behaviour patterns in many online platforms.\\\vspace{-0.12in}

\item Graph-based sequential recommendation methods (\eg SR-GNN, GCSAN, SURGE) perform worse than general sequential baselines (\eg BERT4Rec, HPMN) on IJCAI dataset with longer item sequences. The possible reason is that passing message between items over the generated graph structures based on their directly transitional relations can hardly capture the long-term item dependencies. However, the performance superiority of GNN-based models can be observed on Retailrocket with shorter item sequences. In such cases, modeling of short-term item transitional regularities is sufficient for capturing item dependencies.\\\vspace{-0.12in}


\item BERT4Rec-MB outperforms BERT4Rec in most evaluation cases. In addition, it can be found that multi-behavior recommendation models (\eg MB-GCN, MB-GMN) achieve comparable performance to other baselines. These observations indicate the effectiveness of incorporating multi-behavior context into the learning process of user preference. With the effective modeling of dynamic multi-behaviour patterns from both short-term and long-term perspectives, our \model\ is more effective than those stationary multi-behavior recommendation approaches.
    
\end{itemize}

\begin{table}[t]
\caption{Ablation study with key modules.}
\label{tab:ab}
\vspace{-0.12in}
\resizebox{\linewidth}{!}{
\begin{tabular}{l|cc|cc|cc}
\hline
\multirow{2}{*}{Model Variants} & \multicolumn{2}{c|}{Taobao} & \multicolumn{2}{c|}{Retailrocket} & \multicolumn{2}{c}{IJCAI} \\ \cline{2-7} 
 & HR@5 & NDCG@5 & HR@5 & NDCG@5 & HR@5 & NDCG@5 \\ \hline
\baby & \textbf{0.323}  & \textbf{0.257} & \textbf{0.956} & \textbf{0.950} & \textbf{0.346} & \textbf{0.268} \\
\hline
\textit{(-) MB-Hyper} & 0.261 & 0.206 & 0.883 & 0.861 & 0.320  & 0.249  \\
\textit{(-) ML-Hyper} & 0.271 & 0.212 & 0.898 & 0.874 & 0.328 & 0.256   \\
\textit{(-) Hypergraph} & 0.246 & 0.194 & 0.813 & 0.839 & 0.301 & 0.234  \\
\textit{(-) MS-Attention} & 0.253 & 0.200 & 0.816 & 0.832 & 0.329 & 0.256    \\
 \hline
\end{tabular}
}\vspace{-0.25in}
\end{table}

\vspace{-0.05in}
\subsection{Ablation Study (RQ2)}
\label{exp:ab}
\subsubsection{\rm \textbf{Effects of Key Components.}}
We firstly investigate the effectiveness of different components of our \baby from both Transformer and Hypergraph learning views. Specifically, we generate four variants and make comparison with our \model\ method:
\vspace{-0.05in}
\begin{itemize}[leftmargin=*]

\item \textit{(-) MB-Hyper.} This variant does not include the hypergraph of item-wise multi-behavior dependency to capture the long-range cross-type behavior correlations. 
    

\item \textit{(-) ML-Hyper.} In this variant, we remove the hypergraph message passing over the hyperedges of item semantic dependence (encoded with the metric learning component).
    

\item \textit{(-) Hypergraph.} This variant disables the entire hypergraph item-wise dependency learning, and only relies on the multi-scale Transformer to model the sequential behavior patterns.

\item \textit{(-) MS-Attention.} For this variant, we replace our multi-scale attention layer with the original multi-head attentional operation.

\end{itemize}


From the reported results in Table~\ref{tab:ab}, we summarize the following observations to show the rationality of our model design. 1) With the incorporation of hypergraph-based dependency learning on either item-wise latent semantic ((-) MB-Hyper) or dynamic multi-behavior correlations ((-) ML-Hyper), \model\ can further boost the recommendation performance. 2) Comparing with the vanilla multi-head attention ((-) MS-Attention), the effectiveness of our multi-scale low-rank self-attention can be validated.


\begin{figure}[t]
\centering
\subfigure[Taobao]{
\label{fig:conv:tmall}
\includegraphics[width=0.31\linewidth]{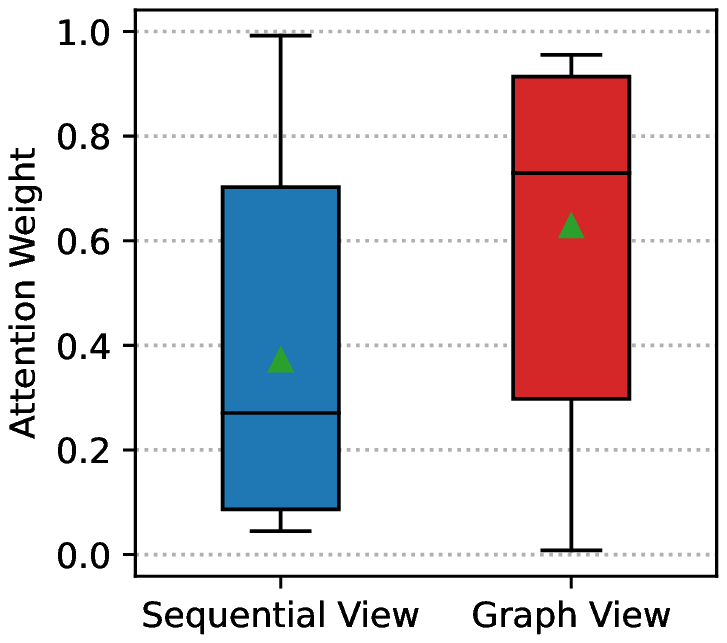}}
\subfigure[IJCAI]{
\label{fig:conv:ijcai}
\includegraphics[width=0.31\linewidth]{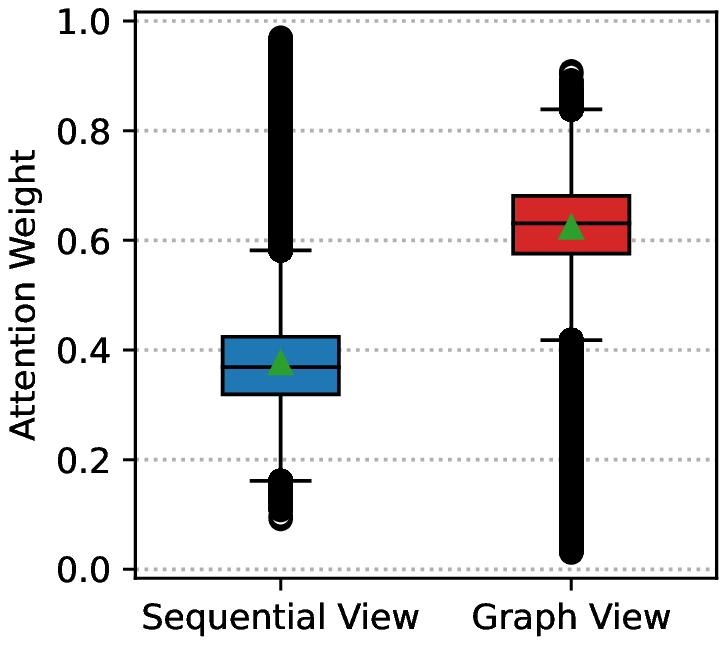}}
\subfigure[Retailrocket]{
\label{fig:conv:ijcai}
\includegraphics[width=0.31\linewidth]{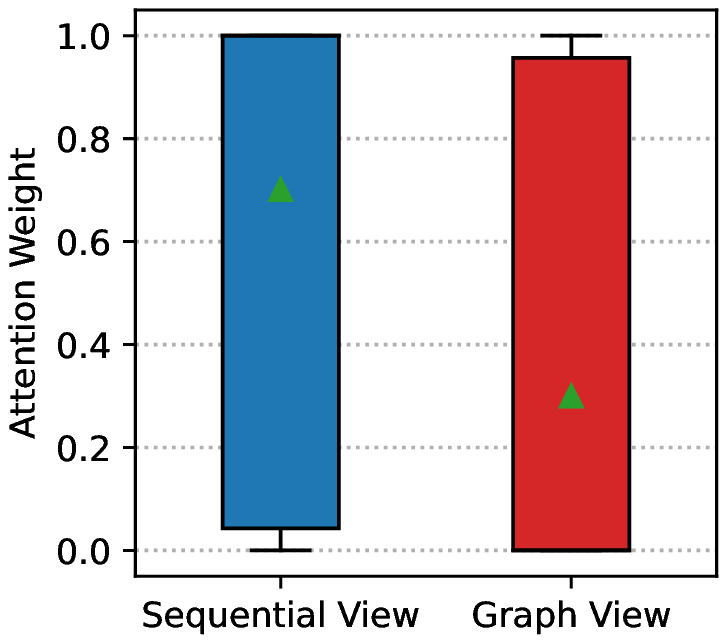}}
\vspace{-0.2in}
\caption{Distributions of the learned attentive view-specific contributions. Green triangles and black line in the showed boxes denote the mean and median values, respectively.}
\label{fig:contrib}
\vspace{-0.2in}
\end{figure}

\subsubsection{\rm \textbf{Contribution of Learning Views.}}
We further investigate the contribution of sequential and hypergraph learning views, by presenting the distributions of our learned importance scores of $\textbf{h}_i$ and $\textbf{x}_i$ in Figure \ref{fig:contrib}. In particular, $\textbf{h}_i$ preserves multi-scale behavior dynamics of diverse user preference and $\textbf{x}_i$ encodes the global multi-behavior dependencies. We can observe that hypergraph learning view contributes more to the effective modeling of dynamic multi-behavior patterns with longer item sequences (\eg Taobao and IJCAI dataset). This further confirms the efficacy of our behavior-aware hypergraph learning component in capturing the long-range item dependencies in multi-relational sequential recommendation.


\begin{figure}[t]
\centering
\subfigure[Taobao]{
\label{fig:group_test:taobao}
\includegraphics[width=0.46\linewidth]{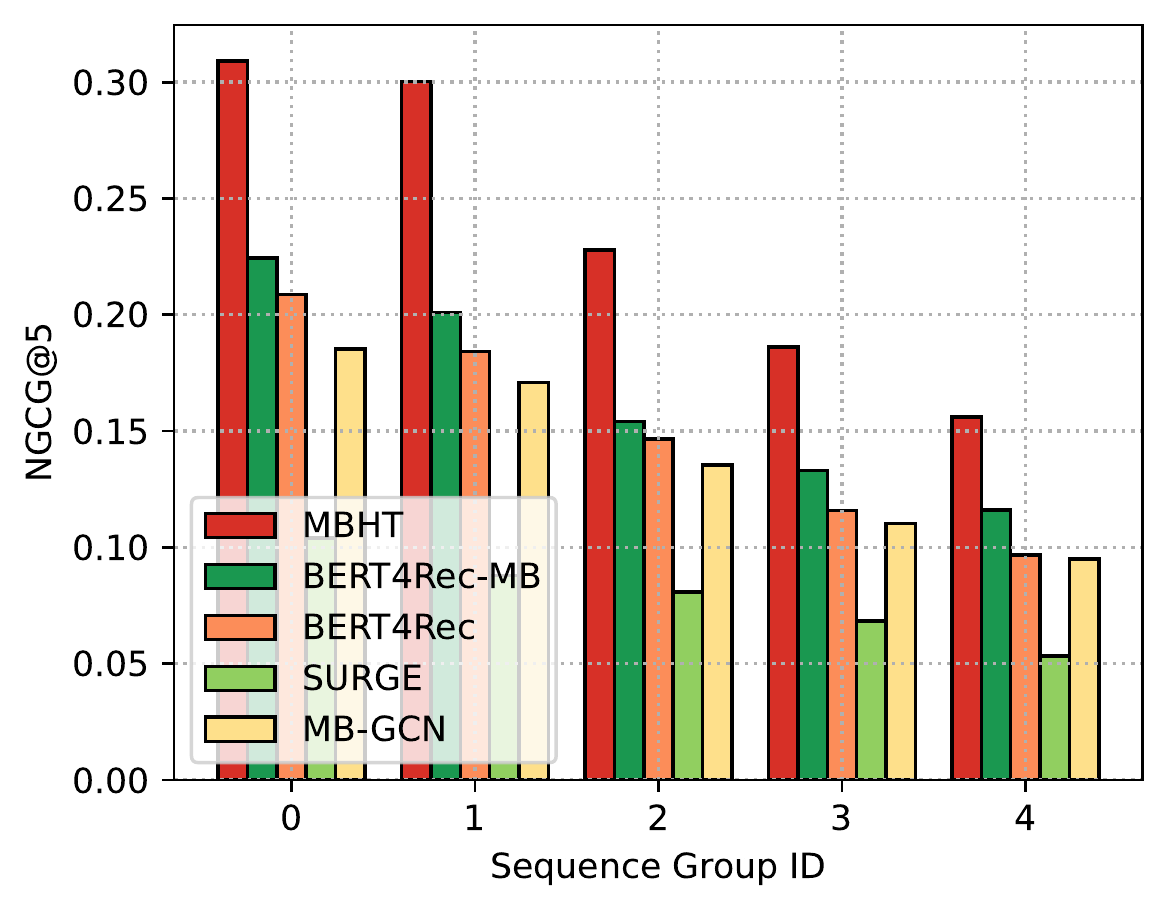}}
\subfigure[IJCAI]{
\label{fig:group_test:ijcai}
\includegraphics[width=0.46\linewidth]{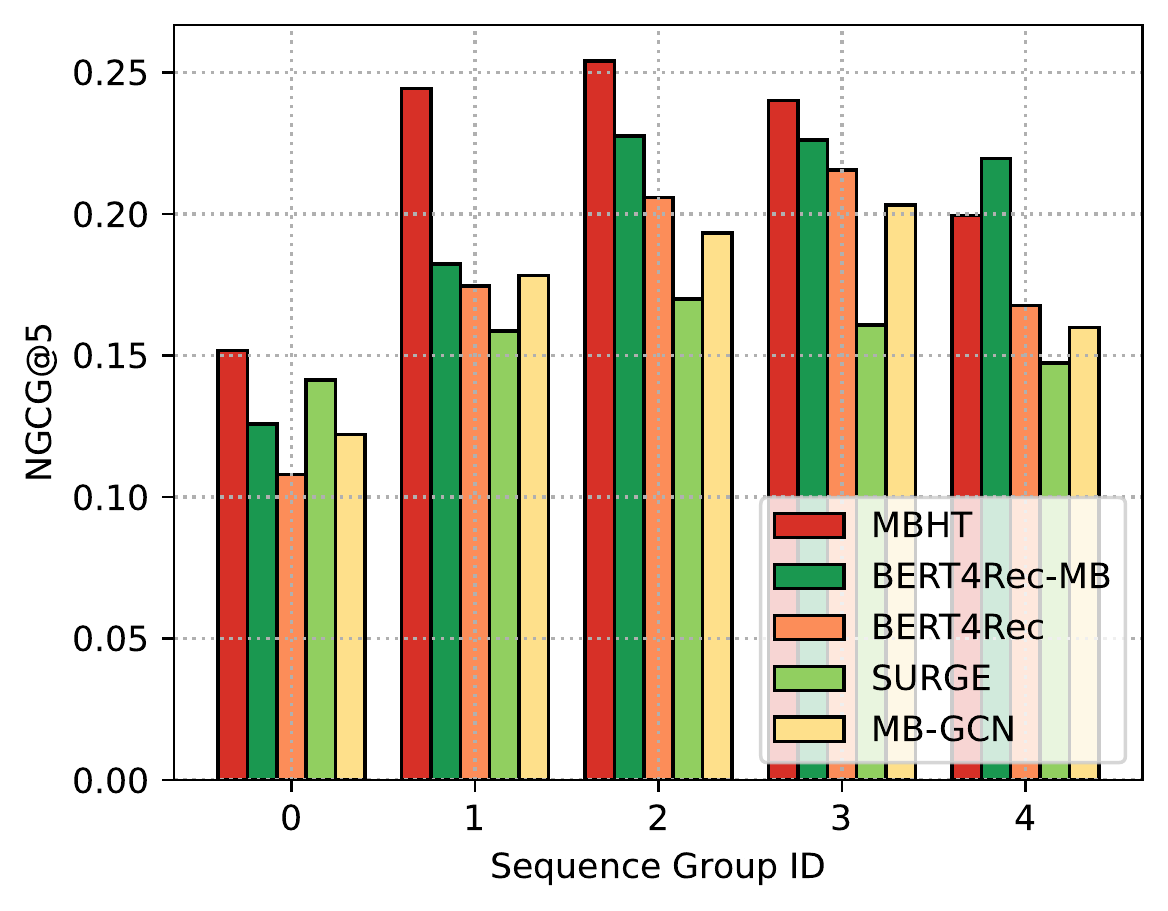}}
\vspace{-0.15in}
\caption{Performance \textit{w.r.t} different sequence lengths.}
\label{fig:group_test}
\vspace{-0.1in}
\end{figure}

\subsection{Model Benefit Study (RQ3)}
\subsubsection{\rm \textbf{Performance \textit{w.r.t} Sequence Length.}}
To further study the robustness of our model, we evaluate \model\ on item sequences with different length. Specifically, we split users into five groups in terms of their item sequences and conduct the performance comparison on each group of users. From results presented in Figure~\ref{fig:group_test}, \model\ outperforms several representative baselines not only on the shorter item interaction sequences, but also on the longer item sequences. It indicates that our recommendation framework is enhanced by injecting the behavior-aware short-term and long-term dependencies (from locally to globally) into the sequence embeddings. Such data scarcity issue is hard to be alleviated purely from the general and GNN-based sequential recommender systems.\vspace{-0.05in}


\begin{figure}[t]
\centering
\subfigure[Taobao]{
\label{fig:conv:tmall}
\includegraphics[width=0.48\linewidth]{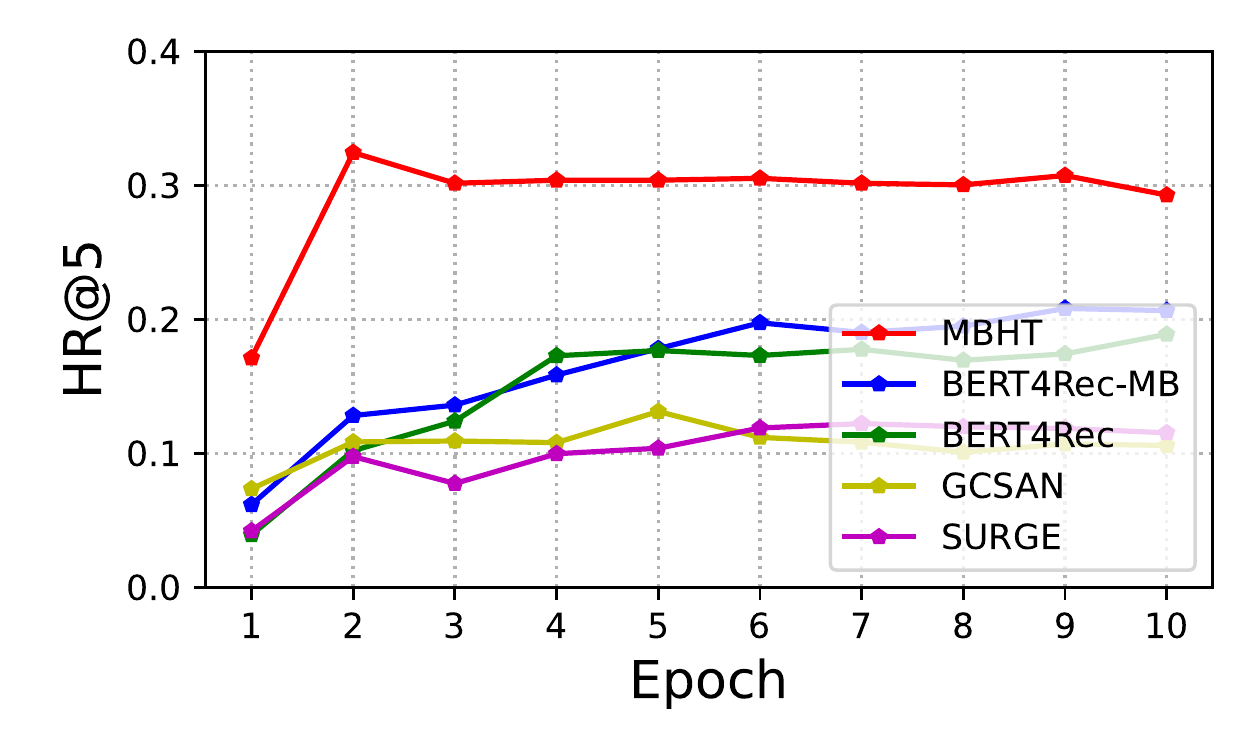}}
\subfigure[IJCAI]{
\label{fig:conv:ijcai}
\includegraphics[width=0.48\linewidth]{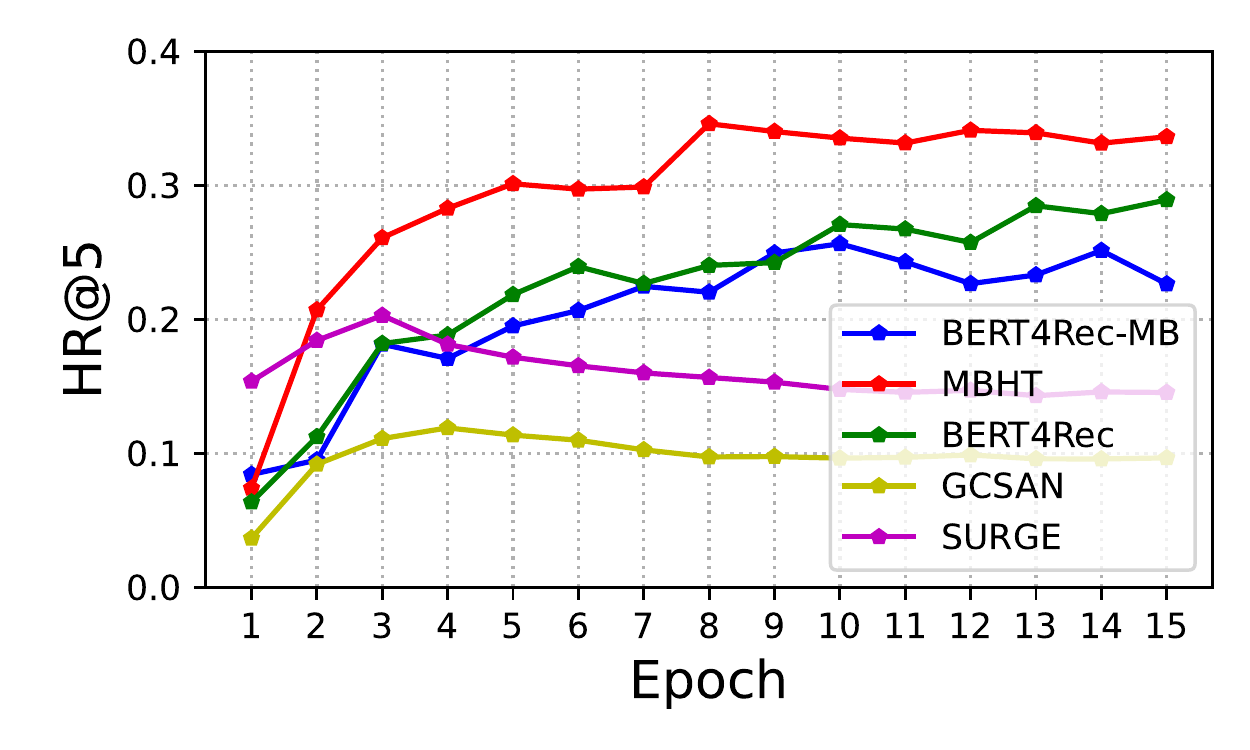}}
\vspace{-0.1in}
\caption{Training curves evaluated by testing Hit Rate.}
\label{fig:convergence}
\vspace{-0.15in}
\end{figure}

\subsubsection{\rm \textbf{Model Convergence Study.}}
We further investigate the convergence property of our \model\ and various sequential and graph-based temporal recommendation methods in Figure~\ref{fig:convergence}. Along the model training process, \model\ achieves faster convergence rate compared with most competitive methods. For example, \model\ obtains its best  performance at epoch 2 and 8, while BERT4Rec and BERT4Rec-MB take 10 and 15 epochs to converge on Taobao and IJCAI datasets, respectively. This observation suggests that exploring the augmented multi-behavior information from both sequential and hypergraph views can provide better gradient to guide the model optimization in sequential recommendation.

\begin{figure}[t]
    \centering
    \includegraphics[width=\linewidth]{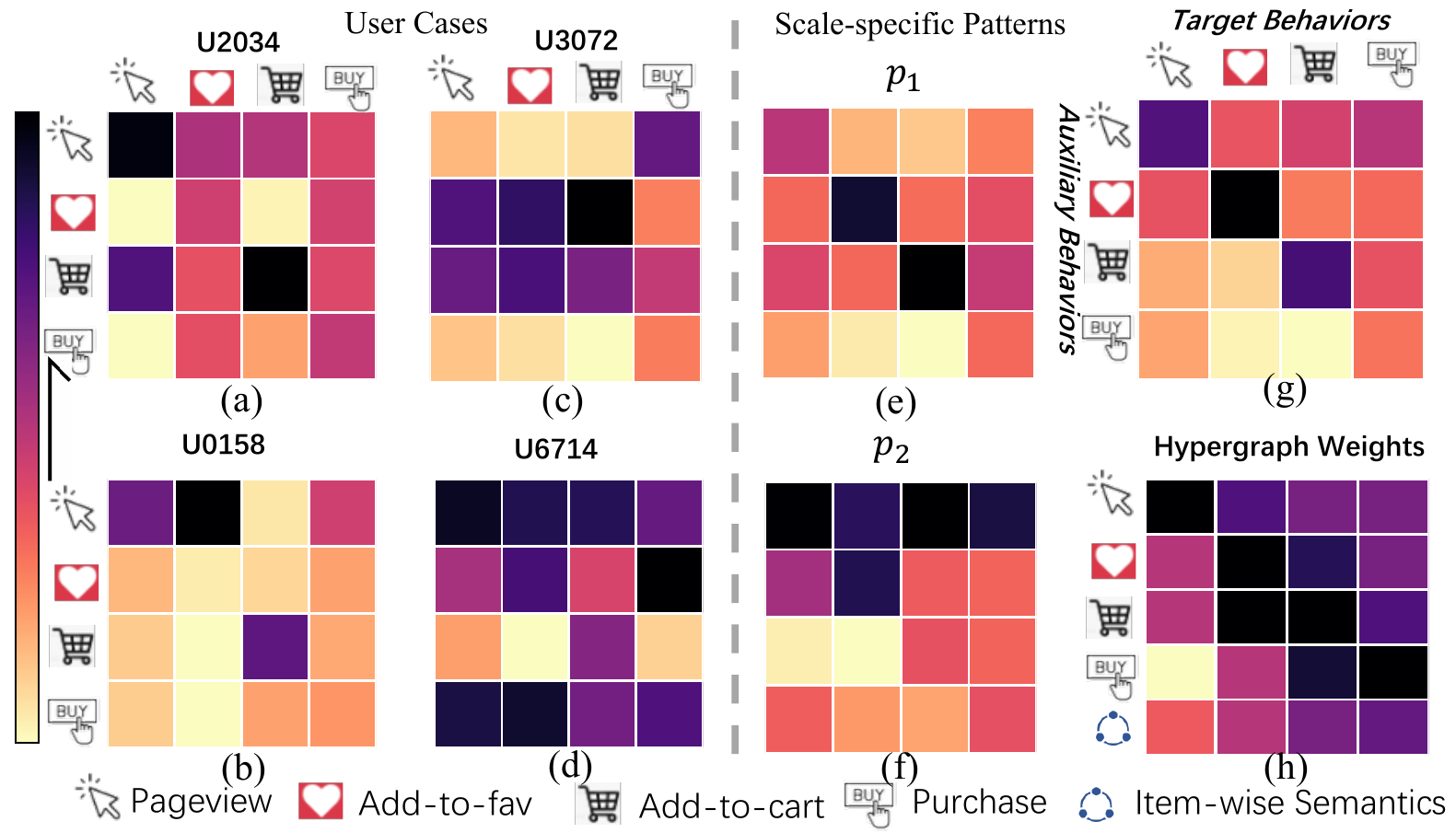}
    \vspace{-0.2in}
    \caption{Case studies with cross-type behavior dependencies.}
    \vspace{-0.2in}
    \label{fig:case_study}
\end{figure}

\vspace{-0.05in}
\subsection{Case Study}
In this section, we conduct further model analysis with case study to show the model interpretation for multi-behavior dependency modeling. In particular, we show user-specific cross-type behavior dependencies in Figure~\ref{fig:case_study} (a)-(d). Each $4\times 4$ dependency matrix is learned from our multi-scale Transformer. We compute the item-item correlations by considering their behavior-aware interactions in the specific sequence $S_i$ of user $u_i$. Figure~\ref{fig:case_study} (e)-(f) show the scale-specific multi-behavior dependencies with the scales of $p_1$ and $p_2$ in our multi-scale modeling of behavior-aware sequential patterns. In Figure~\ref{fig:case_study} (g), we show the overall relevance scores among different types of behaviors in making final forecasting on target behaviors. Additionally, our hypergraph-based item dependencies ($\boldsymbol{\mathcal{M}} \cdot \boldsymbol{\mathcal{M}}^\trans$) are shown in Figure~\ref{fig:case_study} (h), such as hypergraph-based i) behavior-aware item relevance; and ii) item-wise semantic dependence.
\section{Related Work}

\noindent \textbf{Sequential Recommendation}.
Earlier studies solve the next-item recommendation using the Markov Chain-based approaches to model item-item transitions~\cite{rendle2010mc, he2016fusing}. In recent years, many efforts have been devoted to proposing neural network-enhanced sequential recommender systems to encode the complex dependencies among items from different perspectives. For example, recurrent neural network in GRU4Rec~\cite{hidasi2015session} and convolutional operations in Caser~\cite{tang2018caser}.
Inspired by the strength of Transformer, SASRec~\cite{kang2018sasrec} and BERT4Rec~\cite{sun2019bert4rec} are built upon the self-attention mechanism for item-item relation modeling. Furthermore, recently emerged graph neural networks produce state-of-the-art performance by performing graph-based message passing among neighboring items, in order to capture sequential signals, \eg SR-GNN~\cite{wu2019srgnn}, MTD~\cite{huang2021graph}, MA-GNN~\cite{ma2020memory} and SURGE~\cite{chang2021surge}. However, most of those methods are specifically designed for singular type of interaction behaviors, and cannot handle diverse user-item relationships. \\\vspace{-0.12in}

\noindent \textbf{Hypergraph Learning for Recommendation}. Motivated by expressiveness of hypergraphs~\cite{feng2019hypergraph,yi2020hypergraph}, hypergraph neural networks are utilized in several recent recommender systems to model high-order relationships, such as multi-order item correlations in HyperRec~\cite{wang2020next}, global user dependencies in HCCF~\cite{xia2022hypergraph}, high-order social relationships in MHCN~\cite{yu2021self}, and item dependencies with multi-modal features in HyperCTR~\cite{he2021click}. Following this research line, this work integrates the hypergraph neural architecture with Transformer architecture to encode behavior-aware sequential patterns from local to global-level for comprehensive behavior modeling. \\\vspace{-0.12in}

\noindent \textbf{Multi-Behavior Recommender Systems}. There exist recently developed multi-behavior recommender systems for modeling user-item relation heterogeneity~\cite{zhang2020multiplex,gao2019nmtr,jin2020mbgcn,yang2021hyper,wei2022contrastive}. For example, NMTR~\cite{gao2019nmtr} is a multi-task recommendation framework with the predefined behavior-wise cascading relationships. Motivated by the strength of GNNs, MBGCN~\cite{jin2020mbgcn}, MBGMN~\cite{xia2021graph}, MGNN~\cite{zhang2020multiplex} are developed based on the graph-structured message passing over the generated multi-relational user-item interaction graphs. Nevertheless, none of those approaches considers the time-evolving multi-behaviour user preference. To fill this gap, our \model\ model is able to capture both short-term and long-term multi-behavior dependencies with a hypergraph-enhanced transformer architecture.

\section{Conclusion}
In this paper, we present a new sequential recommendation framework \model\ which explicitly captures both short-term and long-term multi-behavior dependencies. \model\ designs a multi-scale Transformer to encode the behavior-aware sequential patterns at both fine-grained and coarse-grained levels. To capture the global cross-type behavior dependencies, we empower \model\ with a multi-behavior hypergraph learning component. Empirical results on several real-world datasets validate the strengths of our \model\ when competing with state-of-the-art recommendation methods.


\section*{Acknowledgments}
This research is supported by the research grants from the Department of Computer Science \& Musketeers Foundation Institute of Data Science at the University of Hong Kong.

\bibliographystyle{ACM-Reference-Format}
\bibliography{reference}


\begin{thebibliography}{36}


\ifx \showCODEN    \undefined \def \showCODEN     #1{\unskip}     \fi
\ifx \showDOI      \undefined \def \showDOI       #1{#1}\fi
\ifx \showISBNx    \undefined \def \showISBNx     #1{\unskip}     \fi
\ifx \showISBNxiii \undefined \def \showISBNxiii  #1{\unskip}     \fi
\ifx \showISSN     \undefined \def \showISSN      #1{\unskip}     \fi
\ifx \showLCCN     \undefined \def \showLCCN      #1{\unskip}     \fi
\ifx \shownote     \undefined \def \shownote      #1{#1}          \fi
\ifx \showarticletitle \undefined \def \showarticletitle #1{#1}   \fi
\ifx \showURL      \undefined \def \showURL       {\relax}        \fi
\providecommand\bibfield[2]{#2}
\providecommand\bibinfo[2]{#2}
\providecommand\natexlab[1]{#1}
\providecommand\showeprint[2][]{arXiv:#2}

\bibitem[\protect\citeauthoryear{Chang, Gao, Zheng, Hui, Niu, Song,
  et~al\mbox{.}}{Chang et~al\mbox{.}}{2021}]%
        {chang2021surge}
\bibfield{author}{\bibinfo{person}{Jianxin Chang}, \bibinfo{person}{Chen Gao},
  \bibinfo{person}{Yu Zheng}, \bibinfo{person}{Yiqun Hui},
  \bibinfo{person}{Yanan Niu}, \bibinfo{person}{Yang Song}, {et~al\mbox{.}}}
  \bibinfo{year}{2021}\natexlab{}.
\newblock \showarticletitle{Sequential Recommendation with Graph Neural
  Networks}. In \bibinfo{booktitle}{\emph{SIGIR}}. \bibinfo{pages}{378--387}.
\newblock


\bibitem[\protect\citeauthoryear{Feng, You, Zhang, Ji, and Gao}{Feng
  et~al\mbox{.}}{2019}]%
        {feng2019hypergraph}
\bibfield{author}{\bibinfo{person}{Yifan Feng}, \bibinfo{person}{Haoxuan You},
  \bibinfo{person}{Zizhao Zhang}, \bibinfo{person}{Rongrong Ji}, {and}
  \bibinfo{person}{Yue Gao}.} \bibinfo{year}{2019}\natexlab{}.
\newblock \showarticletitle{Hypergraph neural networks}. In
  \bibinfo{booktitle}{\emph{AAAI}}, Vol.~\bibinfo{volume}{33}.
  \bibinfo{pages}{3558--3565}.
\newblock


\bibitem[\protect\citeauthoryear{Gao, He, Gan, Chen, et~al\mbox{.}}{Gao
  et~al\mbox{.}}{2019}]%
        {gao2019nmtr}
\bibfield{author}{\bibinfo{person}{Chen Gao}, \bibinfo{person}{Xiangnan He},
  \bibinfo{person}{Dahua Gan}, \bibinfo{person}{Xiangning Chen},
  {et~al\mbox{.}}} \bibinfo{year}{2019}\natexlab{}.
\newblock \showarticletitle{Neural multi-task recommendation from
  multi-behavior data}. In \bibinfo{booktitle}{\emph{ICDE}}. IEEE,
  \bibinfo{pages}{1554--1557}.
\newblock


\bibitem[\protect\citeauthoryear{He, Chen, Wang, Jameel, Yu, et~al\mbox{.}}{He
  et~al\mbox{.}}{2021}]%
        {he2021click}
\bibfield{author}{\bibinfo{person}{Li He}, \bibinfo{person}{Hongxu Chen},
  \bibinfo{person}{Dingxian Wang}, \bibinfo{person}{Shoaib Jameel},
  \bibinfo{person}{Philip Yu}, {et~al\mbox{.}}}
  \bibinfo{year}{2021}\natexlab{}.
\newblock \showarticletitle{Click-Through Rate Prediction with Multi-Modal
  Hypergraphs}. In \bibinfo{booktitle}{\emph{CIKM}}. \bibinfo{pages}{690--699}.
\newblock


\bibitem[\protect\citeauthoryear{He and McAuley}{He and McAuley}{2016}]%
        {he2016fusing}
\bibfield{author}{\bibinfo{person}{Ruining He} {and} \bibinfo{person}{Julian
  McAuley}.} \bibinfo{year}{2016}\natexlab{}.
\newblock \showarticletitle{Fusing similarity models with markov chains for
  sparse sequential recommendation}. In \bibinfo{booktitle}{\emph{ICDM}}. IEEE,
  \bibinfo{pages}{191--200}.
\newblock


\bibitem[\protect\citeauthoryear{He, Deng, Wang, et~al\mbox{.}}{He
  et~al\mbox{.}}{2020}]%
        {he2020lightgcn}
\bibfield{author}{\bibinfo{person}{Xiangnan He}, \bibinfo{person}{Kuan Deng},
  \bibinfo{person}{Xiang Wang}, {et~al\mbox{.}}}
  \bibinfo{year}{2020}\natexlab{}.
\newblock \showarticletitle{Lightgcn: Simplifying and powering graph
  convolution network for recommendation}. In
  \bibinfo{booktitle}{\emph{SIGIR}}. \bibinfo{pages}{639--648}.
\newblock


\bibitem[\protect\citeauthoryear{Hidasi, Karatzoglou, Baltrunas, and
  Tikk}{Hidasi et~al\mbox{.}}{2016}]%
        {hidasi2015session}
\bibfield{author}{\bibinfo{person}{Bal{\'a}zs Hidasi},
  \bibinfo{person}{Alexandros Karatzoglou}, \bibinfo{person}{Linas Baltrunas},
  {and} \bibinfo{person}{Domonkos Tikk}.} \bibinfo{year}{2016}\natexlab{}.
\newblock \showarticletitle{Session-based recommendations with recurrent neural
  networks}. In \bibinfo{booktitle}{\emph{ICLR}}.
\newblock


\bibitem[\protect\citeauthoryear{Huang, Chen, Xia, Xu, Dai, Chen, Bo, Zhao, and
  Huang}{Huang et~al\mbox{.}}{2021}]%
        {huang2021graph}
\bibfield{author}{\bibinfo{person}{Chao Huang}, \bibinfo{person}{Jiahui Chen},
  \bibinfo{person}{Lianghao Xia}, \bibinfo{person}{Yong Xu},
  \bibinfo{person}{Peng Dai}, \bibinfo{person}{Yanqing Chen},
  \bibinfo{person}{Liefeng Bo}, \bibinfo{person}{Jiashu Zhao}, {and}
  \bibinfo{person}{Jimmy~Xiangji Huang}.} \bibinfo{year}{2021}\natexlab{}.
\newblock \showarticletitle{Graph-enhanced multi-task learning of multi-level
  transition dynamics for session-based recommendation}. In
  \bibinfo{booktitle}{\emph{AAAI}}.
\newblock


\bibitem[\protect\citeauthoryear{Huang, Wu, Zhang, Zhang, Zhao, Yin, and
  Chawla}{Huang et~al\mbox{.}}{2019}]%
        {2019online}
\bibfield{author}{\bibinfo{person}{Chao Huang}, \bibinfo{person}{Xian Wu},
  \bibinfo{person}{Xuchao Zhang}, \bibinfo{person}{Chuxu Zhang},
  \bibinfo{person}{Jiashu Zhao}, \bibinfo{person}{Dawei Yin}, {and}
  \bibinfo{person}{Nitesh~V Chawla}.} \bibinfo{year}{2019}\natexlab{}.
\newblock \showarticletitle{Online purchase prediction via multi-scale modeling
  of behavior dynamics}. In \bibinfo{booktitle}{\emph{KDD}}.
  \bibinfo{pages}{2613--2622}.
\newblock


\bibitem[\protect\citeauthoryear{Jiang et~al\mbox{.}}{Jiang
  et~al\mbox{.}}{2020}]%
        {jiang2020aspect}
\bibfield{author}{\bibinfo{person}{Hao Jiang} {et~al\mbox{.}}}
  \bibinfo{year}{2020}\natexlab{}.
\newblock \showarticletitle{What aspect do you like: Multi-scale time-aware
  user interest modeling for micro-video recommendation}. In
  \bibinfo{booktitle}{\emph{MM}}. \bibinfo{pages}{3487--3495}.
\newblock


\bibitem[\protect\citeauthoryear{Jin, Gao, He, Jin, et~al\mbox{.}}{Jin
  et~al\mbox{.}}{2020}]%
        {jin2020mbgcn}
\bibfield{author}{\bibinfo{person}{Bowen Jin}, \bibinfo{person}{Chen Gao},
  \bibinfo{person}{Xiangnan He}, \bibinfo{person}{Depeng Jin}, {et~al\mbox{.}}}
  \bibinfo{year}{2020}\natexlab{}.
\newblock \showarticletitle{Multi-behavior recommendation with graph
  convolutional networks}. In \bibinfo{booktitle}{\emph{SIGIR}}.
  \bibinfo{pages}{659--668}.
\newblock


\bibitem[\protect\citeauthoryear{Kang, Lee, Choe, and Jung}{Kang
  et~al\mbox{.}}{2021}]%
        {kang2021entangled}
\bibfield{author}{\bibinfo{person}{Taegwan Kang}, \bibinfo{person}{Hwanhee
  Lee}, \bibinfo{person}{Byeongjin Choe}, {and} \bibinfo{person}{Kyomin Jung}.}
  \bibinfo{year}{2021}\natexlab{}.
\newblock \showarticletitle{Entangled bidirectional encoder to autoregressive
  decoder for sequential recommendation}. In \bibinfo{booktitle}{\emph{SIGIR}}.
  \bibinfo{pages}{1657--1661}.
\newblock


\bibitem[\protect\citeauthoryear{Kang and McAuley}{Kang and McAuley}{2018}]%
        {kang2018sasrec}
\bibfield{author}{\bibinfo{person}{Wang-Cheng Kang} {and}
  \bibinfo{person}{Julian McAuley}.} \bibinfo{year}{2018}\natexlab{}.
\newblock \showarticletitle{Self-attentive sequential recommendation}. In
  \bibinfo{booktitle}{\emph{ICDM}}. IEEE, \bibinfo{pages}{197--206}.
\newblock


\bibitem[\protect\citeauthoryear{Kitaev et~al\mbox{.}}{Kitaev
  et~al\mbox{.}}{2020}]%
        {kitaev2020reformer}
\bibfield{author}{\bibinfo{person}{Nikita Kitaev} {et~al\mbox{.}}}
  \bibinfo{year}{2020}\natexlab{}.
\newblock \showarticletitle{Reformer: The efficient transformer}. In
  \bibinfo{booktitle}{\emph{ICLR}}.
\newblock


\bibitem[\protect\citeauthoryear{Liu, Lin, Cao, Hu, Wei, et~al\mbox{.}}{Liu
  et~al\mbox{.}}{2021}]%
        {liu2021swin}
\bibfield{author}{\bibinfo{person}{Ze Liu}, \bibinfo{person}{Yutong Lin},
  \bibinfo{person}{Yue Cao}, \bibinfo{person}{Han Hu}, \bibinfo{person}{Yixuan
  Wei}, {et~al\mbox{.}}} \bibinfo{year}{2021}\natexlab{}.
\newblock \showarticletitle{Swin transformer: Hierarchical vision transformer
  using shifted windows}. In \bibinfo{booktitle}{\emph{ICCV}}.
  \bibinfo{pages}{10012--10022}.
\newblock


\bibitem[\protect\citeauthoryear{Ma, Ma, Zhang, et~al\mbox{.}}{Ma
  et~al\mbox{.}}{2020}]%
        {ma2020memory}
\bibfield{author}{\bibinfo{person}{Chen Ma}, \bibinfo{person}{Liheng Ma},
  \bibinfo{person}{Yingxue Zhang}, {et~al\mbox{.}}}
  \bibinfo{year}{2020}\natexlab{}.
\newblock \showarticletitle{Memory augmented graph neural networks for
  sequential recommendation}. In \bibinfo{booktitle}{\emph{AAAI}},
  Vol.~\bibinfo{volume}{34}. \bibinfo{pages}{5045--5052}.
\newblock


\bibitem[\protect\citeauthoryear{Ren, Qin, et~al\mbox{.}}{Ren
  et~al\mbox{.}}{2019}]%
        {ren2019lifelong}
\bibfield{author}{\bibinfo{person}{Kan Ren}, \bibinfo{person}{Jiarui Qin},
  {et~al\mbox{.}}} \bibinfo{year}{2019}\natexlab{}.
\newblock \showarticletitle{Lifelong sequential modeling with personalized
  memorization for user response prediction}. In
  \bibinfo{booktitle}{\emph{SIGIR}}. \bibinfo{pages}{565--574}.
\newblock


\bibitem[\protect\citeauthoryear{Rendle, Freudenthaler, et~al\mbox{.}}{Rendle
  et~al\mbox{.}}{2010}]%
        {rendle2010mc}
\bibfield{author}{\bibinfo{person}{Steffen Rendle}, \bibinfo{person}{Christoph
  Freudenthaler}, {et~al\mbox{.}}} \bibinfo{year}{2010}\natexlab{}.
\newblock \showarticletitle{Factorizing personalized markov chains for
  next-basket recommendation}. In \bibinfo{booktitle}{\emph{WWW}}.
  \bibinfo{pages}{811--820}.
\newblock


\bibitem[\protect\citeauthoryear{Sun, Liu, et~al\mbox{.}}{Sun
  et~al\mbox{.}}{2019}]%
        {sun2019bert4rec}
\bibfield{author}{\bibinfo{person}{Fei Sun}, \bibinfo{person}{Jun Liu},
  {et~al\mbox{.}}} \bibinfo{year}{2019}\natexlab{}.
\newblock \showarticletitle{BERT4Rec: Sequential recommendation with
  bidirectional encoder representations from transformer}. In
  \bibinfo{booktitle}{\emph{CIKM}}. \bibinfo{pages}{1441--1450}.
\newblock


\bibitem[\protect\citeauthoryear{Tang and Wang}{Tang and Wang}{2018}]%
        {tang2018caser}
\bibfield{author}{\bibinfo{person}{Jiaxi Tang} {and} \bibinfo{person}{Ke
  Wang}.} \bibinfo{year}{2018}\natexlab{}.
\newblock \showarticletitle{Personalized top-n sequential recommendation via
  convolutional sequence embedding}. In \bibinfo{booktitle}{\emph{WSDM}}.
  \bibinfo{pages}{565--573}.
\newblock


\bibitem[\protect\citeauthoryear{Wang, Ding, Hong, Liu, and Caverlee}{Wang
  et~al\mbox{.}}{2020a}]%
        {wang2020next}
\bibfield{author}{\bibinfo{person}{Jianling Wang}, \bibinfo{person}{Kaize
  Ding}, \bibinfo{person}{Liangjie Hong}, \bibinfo{person}{Huan Liu}, {and}
  \bibinfo{person}{James Caverlee}.} \bibinfo{year}{2020}\natexlab{a}.
\newblock \showarticletitle{Next-item recommendation with sequential
  hypergraphs}. In \bibinfo{booktitle}{\emph{SIGIR}}.
  \bibinfo{pages}{1101--1110}.
\newblock


\bibitem[\protect\citeauthoryear{Wang, Li, Khabsa, Fang, and Ma}{Wang
  et~al\mbox{.}}{2020b}]%
        {wang2020linformer}
\bibfield{author}{\bibinfo{person}{Sinong Wang}, \bibinfo{person}{Belinda~Z.
  Li}, \bibinfo{person}{Madian Khabsa}, \bibinfo{person}{Han Fang}, {and}
  \bibinfo{person}{Hao Ma}.} \bibinfo{year}{2020}\natexlab{b}.
\newblock \bibinfo{title}{Linformer: Self-Attention with Linear Complexity}.
\newblock
\newblock
\showeprint[arxiv]{2006.04768}~[cs.LG]


\bibitem[\protect\citeauthoryear{Wei, Huang, Xia, Xu, Zhao, and Yin}{Wei
  et~al\mbox{.}}{2022}]%
        {wei2022contrastive}
\bibfield{author}{\bibinfo{person}{Wei Wei}, \bibinfo{person}{Chao Huang},
  \bibinfo{person}{Lianghao Xia}, \bibinfo{person}{Yong Xu},
  \bibinfo{person}{Jiashu Zhao}, {and} \bibinfo{person}{Dawei Yin}.}
  \bibinfo{year}{2022}\natexlab{}.
\newblock \showarticletitle{Contrastive Meta Learning with Behavior
  Multiplicity for Recommendation}. In \bibinfo{booktitle}{\emph{WSDM}}.
  \bibinfo{pages}{1120--1128}.
\newblock


\bibitem[\protect\citeauthoryear{Wu, Hu, Hong, et~al\mbox{.}}{Wu
  et~al\mbox{.}}{2018}]%
        {wu2018turning}
\bibfield{author}{\bibinfo{person}{Liang Wu}, \bibinfo{person}{Diane Hu},
  \bibinfo{person}{Liangjie Hong}, {et~al\mbox{.}}}
  \bibinfo{year}{2018}\natexlab{}.
\newblock \showarticletitle{Turning clicks into purchases: Revenue optimization
  for product search in e-commerce}. In \bibinfo{booktitle}{\emph{SIGIR}}.
  \bibinfo{pages}{365--374}.
\newblock


\bibitem[\protect\citeauthoryear{Wu, Tang, Zhu, et~al\mbox{.}}{Wu
  et~al\mbox{.}}{2019}]%
        {wu2019srgnn}
\bibfield{author}{\bibinfo{person}{Shu Wu}, \bibinfo{person}{Yuyuan Tang},
  \bibinfo{person}{Yanqiao Zhu}, {et~al\mbox{.}}}
  \bibinfo{year}{2019}\natexlab{}.
\newblock \showarticletitle{Session-based recommendation with graph neural
  networks}. In \bibinfo{booktitle}{\emph{AAAI}}, Vol.~\bibinfo{volume}{33}.
  \bibinfo{pages}{346--353}.
\newblock


\bibitem[\protect\citeauthoryear{Xia, Huang, Xu, Zhao, Yin, and Huang}{Xia
  et~al\mbox{.}}{2022}]%
        {xia2022hypergraph}
\bibfield{author}{\bibinfo{person}{Lianghao Xia}, \bibinfo{person}{Chao Huang},
  \bibinfo{person}{Yong Xu}, \bibinfo{person}{Jiashu Zhao},
  \bibinfo{person}{Dawei Yin}, {and} \bibinfo{person}{Jimmy~Xiangji Huang}.}
  \bibinfo{year}{2022}\natexlab{}.
\newblock \showarticletitle{Hypergraph Contrastive Collaborative Filtering}.
\newblock \bibinfo{journal}{\emph{arXiv preprint arXiv:2204.12200}}.
\newblock


\bibitem[\protect\citeauthoryear{Xia, Xu, Huang, Dai, and Bo}{Xia
  et~al\mbox{.}}{2021}]%
        {xia2021graph}
\bibfield{author}{\bibinfo{person}{Lianghao Xia}, \bibinfo{person}{Yong Xu},
  \bibinfo{person}{Chao Huang}, \bibinfo{person}{Peng Dai}, {and}
  \bibinfo{person}{Liefeng Bo}.} \bibinfo{year}{2021}\natexlab{}.
\newblock \showarticletitle{Graph meta network for multi-behavior
  recommendation}. In \bibinfo{booktitle}{\emph{SIGIR}}.
  \bibinfo{pages}{757--766}.
\newblock


\bibitem[\protect\citeauthoryear{Xu, Zhao, et~al\mbox{.}}{Xu
  et~al\mbox{.}}{2019}]%
        {xu2019gcsan}
\bibfield{author}{\bibinfo{person}{Chengfeng Xu}, \bibinfo{person}{Pengpeng
  Zhao}, {et~al\mbox{.}}} \bibinfo{year}{2019}\natexlab{}.
\newblock \showarticletitle{Graph Contextualized Self-Attention Network for
  Session-based Recommendation.}. In \bibinfo{booktitle}{\emph{IJCAI}},
  Vol.~\bibinfo{volume}{19}. \bibinfo{pages}{3940--3946}.
\newblock


\bibitem[\protect\citeauthoryear{Yang, Chen, Li, Philip, and Xu}{Yang
  et~al\mbox{.}}{2021}]%
        {yang2021hyper}
\bibfield{author}{\bibinfo{person}{Haoran Yang}, \bibinfo{person}{Hongxu Chen},
  \bibinfo{person}{Lin Li}, \bibinfo{person}{S~Yu Philip}, {and}
  \bibinfo{person}{Guandong Xu}.} \bibinfo{year}{2021}\natexlab{}.
\newblock \showarticletitle{Hyper Meta-Path Contrastive Learning for
  Multi-Behavior Recommendation}. In \bibinfo{booktitle}{\emph{ICDM}}. IEEE,
  \bibinfo{pages}{787--796}.
\newblock


\bibitem[\protect\citeauthoryear{Yang, Huang, Xia, and Li}{Yang
  et~al\mbox{.}}{2022}]%
        {yang2022knowledge}
\bibfield{author}{\bibinfo{person}{Yuhao Yang}, \bibinfo{person}{Chao Huang},
  \bibinfo{person}{Lianghao Xia}, {and} \bibinfo{person}{Chenliang Li}.}
  \bibinfo{year}{2022}\natexlab{}.
\newblock \showarticletitle{Knowledge Graph Contrastive Learning for
  Recommendation}.
\newblock \bibinfo{journal}{\emph{arXiv preprint arXiv:2205.00976}}
  (\bibinfo{year}{2022}).
\newblock


\bibitem[\protect\citeauthoryear{Yao and Wan}{Yao and Wan}{2020}]%
        {yao2020multimodal}
\bibfield{author}{\bibinfo{person}{Shaowei Yao} {and} \bibinfo{person}{Xiaojun
  Wan}.} \bibinfo{year}{2020}\natexlab{}.
\newblock \showarticletitle{Multimodal transformer for multimodal machine
  translation}. In \bibinfo{booktitle}{\emph{ACL}}.
  \bibinfo{pages}{4346--4350}.
\newblock


\bibitem[\protect\citeauthoryear{Yi and Park}{Yi and Park}{2020}]%
        {yi2020hypergraph}
\bibfield{author}{\bibinfo{person}{Jaehyuk Yi} {and} \bibinfo{person}{Jinkyoo
  Park}.} \bibinfo{year}{2020}\natexlab{}.
\newblock \showarticletitle{Hypergraph convolutional recurrent neural network}.
  In \bibinfo{booktitle}{\emph{KDD}}. \bibinfo{pages}{3366--3376}.
\newblock


\bibitem[\protect\citeauthoryear{Yu, Yin, et~al\mbox{.}}{Yu
  et~al\mbox{.}}{2021}]%
        {yu2021self}
\bibfield{author}{\bibinfo{person}{Junliang Yu}, \bibinfo{person}{Hongzhi Yin},
  {et~al\mbox{.}}} \bibinfo{year}{2021}\natexlab{}.
\newblock \showarticletitle{Self-Supervised Multi-Channel Hypergraph
  Convolutional Network for Social Recommendation}. In
  \bibinfo{booktitle}{\emph{WWW}}. \bibinfo{pages}{413--424}.
\newblock


\bibitem[\protect\citeauthoryear{Zhang, Yao, Sun, and Tay}{Zhang
  et~al\mbox{.}}{2019}]%
        {zhang2019deep}
\bibfield{author}{\bibinfo{person}{Shuai Zhang}, \bibinfo{person}{Lina Yao},
  \bibinfo{person}{Aixin Sun}, {and} \bibinfo{person}{Yi Tay}.}
  \bibinfo{year}{2019}\natexlab{}.
\newblock \showarticletitle{Deep learning based recommender system: A survey
  and new perspectives}.
\newblock \bibinfo{journal}{\emph{ACM Computing Surveys (CSUR)}}
  \bibinfo{volume}{52}, \bibinfo{number}{1} (\bibinfo{year}{2019}),
  \bibinfo{pages}{1--38}.
\newblock


\bibitem[\protect\citeauthoryear{Zhang, Mao, Cao, and Xu}{Zhang
  et~al\mbox{.}}{2020}]%
        {zhang2020multiplex}
\bibfield{author}{\bibinfo{person}{Weifeng Zhang}, \bibinfo{person}{Jingwen
  Mao}, \bibinfo{person}{Yi Cao}, {and} \bibinfo{person}{Congfu Xu}.}
  \bibinfo{year}{2020}\natexlab{}.
\newblock \showarticletitle{Multiplex graph neural networks for multi-behavior
  recommendation}. In \bibinfo{booktitle}{\emph{CIKM}}.
  \bibinfo{pages}{2313--2316}.
\newblock


\bibitem[\protect\citeauthoryear{Zhou, Qin, Lu, Chen, and Zhang}{Zhou
  et~al\mbox{.}}{2019}]%
        {zhou2019online}
\bibfield{author}{\bibinfo{person}{Xiangmin Zhou}, \bibinfo{person}{Dong Qin},
  \bibinfo{person}{Xiaolu Lu}, \bibinfo{person}{Lei Chen}, {and}
  \bibinfo{person}{Yanchun Zhang}.} \bibinfo{year}{2019}\natexlab{}.
\newblock \showarticletitle{Online social media recommendation over streams}.
  In \bibinfo{booktitle}{\emph{ICDE}}. IEEE, \bibinfo{pages}{938--949}.
\newblock


\end{thebibliography}

\clearpage
\appendix \section{Supplementary Material}
\balance
\label{sec:appendix}
In our supplemental material, we first summarize the learning process of our \model\ framework in Algorithm \ref{algorithm} and conduct the model time complexity analysis. Then, we present our strategy to simplify the implementation of our hypergraph-based embedding propagation, so as to improve the model efficiency.

\subsection{The Learning Process of \model}
\begin{algorithm}
\SetKwInOut{Input}{Input}\SetKwInOut{Output}{Output}
\caption{The forward propagation flow of \baby}
\label{algorithm}
\Input{The behavior-aware interaction sequence $S_i$ for user $u_i$ with mask tokens at positions $M$ and with true labels $T$. $S_i=[(v_{i,1}, b_{i,1}),...,\textsf{[mask]},...,(v_{i,J}, b_{i,J})]$.}
\Output{The estimated probability for user $u_i$ interacting with ground-truth items $T$ at the time step positions $M$.}
\emph{\textbf{Multi-Scale Transformer View}}\;
Inject positional and behavior signals into Transformer inputs: $\textbf{H} \leftarrow [\textbf{h}_1,...,\textbf{h}_{J+1}]$, $\textbf{h}_j \leftarrow \textbf{e}_j \oplus \textbf{p}_j \oplus \textbf{b}_j$\;
Perform multi-scale attention under multi-scale settings $C, p_1$ and $p_2$ according to Equation \ref{eq:lowrank}-\ref{eq:pool_attn}: $\widetilde{\textbf{H}} \leftarrow f(\widehat{\textbf{H}} \mathbin\Vert {\textbf{H}}^{p_1} \mathbin\Vert {\textbf{H}}^{p_2})$\;
Perform point-wise feed-forward to inject non-linearity: $	\widetilde{\textbf{H}}^{(l)} \leftarrow [\textsf{FFN}(\tilde{\textbf{h}}_1^{(l)})^\trans, \cdots, \textsf{FFN}(\tilde{\textbf{h}}_t^{(l)})^\trans]$\;
\emph{\textbf{Hypergraph View}}\;
Model item-wise semantic dependencies with multi-channel metric learning: $	\beta_{j,j'} \leftarrow \frac{1}{N}\sum_{n=1}^{N} \hat{\beta}^n_{j,j'}$\;
Construct customized hyperedges according to Equation \ref{eq:sim_edge}-\ref{eq:mb_edge}: $\mathcal{M} \leftarrow \mathcal{M}^p \mathbin\Vert \mathcal{M}^q$\;
Apply hypergraph convolutional function to aggregate information from the graph: $\textbf{X}^{(l+1)} \leftarrow \textbf{D}_v^{-1} \cdot \boldsymbol{\mathcal{M}} \cdot \textbf{D}_e^{-1} \cdot \boldsymbol{\mathcal{M}}^\trans \cdot \textbf{X}^{(l)}$\;
\emph{\textbf{Fusion and Prediction}}\;
\ForEach{$m \in M, t \in T$}{
    Generate hypergraph-view embedding for mask position $m$ using sliding-window contextual pooling: $\tilde{\textbf{x}}_{m} \leftarrow \textsf{mean}(\tilde{\textbf{x}}_{m-q_1},...,\tilde{\textbf{x}}_{m+q_2})$\;
    Apply attentive cross-view aggregation to fuse the information according to Equation \ref{eq:fuse}: $\textbf{g}_{m} \leftarrow \alpha_1 \cdot \tilde{\textbf{h}}_{m} \oplus \alpha_2 \cdot \tilde{\textbf{x}}_{m}$\;
    Calculate the probability of item at $m$ being $v_t$: $\hat{y}_{m,t} \leftarrow \textbf{g}_{m}^\trans \textbf{v}_{t}$\;
}
\Return{$[\hat{y}_{m_1,t_1}, \hat{y}_{m_2,t_2},...]$}\;
\end{algorithm}

\subsection{Time Complexity Analysis}
\label{sec:time_comp}
This section conducts the time complexity analysis of our \model\ framework as follows. (1) For the multi-scale Transformer view, with our low-rank self-attention layer, we significantly reduce the computational cost from $O(L\times d\times ((\frac{J}{C})^2+(\frac{J}{p_1})^2+(\frac{J}{p_2})^2))$ to approximate the linear time complexity $O(3LdJ)$, given that $C$ (low-rank scale), $p_1,p_2$ (resolution scales) $\ll J$~\cite{wang2020linformer}. $J$ denotes the length of item sequence. (2) For the hypergraph learning view, the semantic metric learning takes $O(J^2d)$ complexity, and the hypergraph convolutional function takes $O(LJd^2)$. 
Based on the above discussion, the overall time complexity of our \model\ is $O(3LJd + LJd^2 + J^2d)$, which is comparable to state-of-the-art baselines.

\subsection{Simplifying Hypergraph Message Passing}
\label{sec:lighthgcn}
With the consideration of high computational cost during the message passing in our hypergraph learning framework, we propose to simplify the propagation scheme with the learnable embedding projection. Formally, we rewrite the two-stage (node-hyperedge-node) message propagation process $\boldsymbol{\mathcal{M}} \cdot \boldsymbol{\mathcal{M}}^\trans$ as follows:
\begin{align}
    (\boldsymbol{\mathcal{M}}\cdot \boldsymbol{\mathcal{M}}^\trans)_{ij} = \sum_{k=1}^n \boldsymbol{\mathcal{M}}_{ik}\boldsymbol{\mathcal{M}}^\trans_{kj} = \boldsymbol{\mathcal{M}}_{(i)}\boldsymbol{\mathcal{M}}^{{\trans}^{(j)}} = \boldsymbol{\mathcal{M}}_{(i)}\boldsymbol{\mathcal{M}}_{(j)}
\end{align}
where $\boldsymbol{\mathcal{M}}_{(i)}$ and $\boldsymbol{\mathcal{M}_{(j)}}$ denotes the $i$-th row and $j$-th column vector, respectively. Based on our item-wise semantic dependence and multi-behavior correlations, our hypergraph-guided information propagation can be presented as follows:
\begin{align}
\label{eq:lhgcn_split}
    \boldsymbol{\mathcal{M}}_{(i)}\boldsymbol{\mathcal{M}}_{(j)} = \sum_{k=1}^{|\mathcal{E}^q|} \boldsymbol{\mathcal{M}}_{i,k}\boldsymbol{\mathcal{M}}_{j,k} + \sum_{m=1}^{|\mathcal{E}^p|} \boldsymbol{\mathcal{M}}_{i,m}\boldsymbol{\mathcal{M}}_{j,m}
\end{align}
\noindent where $\boldsymbol{\mathcal{M}}_{i,k}$ and $\boldsymbol{\mathcal{M}}_{i,m}$ denote the value of hypergraph connection matrix with item-wise semantic dependency and multi-behavior correlations of item $v_i$, respectively. Here, we first simplify the multiplication operations with the multi-behavior dependency hyperedges. Specifically, the same item ($v_i=v_j$) interacted with different behavior types over time will be connected to the same hyperedge $\epsilon^q$, \ie $\boldsymbol{\mathcal{M}}_{i,\epsilon^q}= \boldsymbol{\mathcal{M}}_{j,\epsilon^q} = 1$. Since each item can only be connected to one hyperedge, for $\epsilon' \neq \epsilon^q$, $\boldsymbol{\mathcal{M}}_{i, \epsilon'}= \boldsymbol{\mathcal{M}}_{j, \epsilon^q} = 0$. This indicates that if $v_i \neq v_j$, there exists no multi-behavior dependency hyperedge $\epsilon'$ such that $\boldsymbol{\mathcal{M}}_{i, \epsilon'} \boldsymbol{\mathcal{M}}_{j, \epsilon'}=1$, since different items cannot be connected to the same hyperedge. Hence, we can isolate the influence of the multi-behavior dependency hyperedges as follows:
\begin{align}
    \sum_{\epsilon'=1}^{|\mathcal{E}^q|} \boldsymbol{\mathcal{M}}_{i, \epsilon'}\boldsymbol{\mathcal{M}}_{j, \epsilon'} = \begin{cases}
    1 & \text{if}~~ v_i=v_j \\
    0 & \text{otherwise}
    \end{cases}
\end{align}

We further investigate the item-wise semantic dependency hyperedges in Equation \ref{eq:lhgcn_split} for simplifying hypergraph convolutional operations. Note that, $\boldsymbol{\mathcal{M}}_{i, m}$ denotes the cosine similarity between item $v_i$ and $v_m$ that belong to the hyperedge $m$. Generally, the computation $\sum_{m=1}^{|\mathcal{E}^p|} \boldsymbol{\mathcal{M}}_{im}\boldsymbol{\mathcal{M}}_{jm}$ can be divided into three cases depending on the hyperedge $m$: i) behavior-aware self-connection, if $v_i = v_j$ and $m$ is the hyperedge assigned to this item; ii) first-order similarity, if $v_i \neq v_j$ and $m$ is assigned to $v_i$ or $v_j$; iii) second-order similarity, if $v_i \neq v_j$ and $m$ is not assigned to either $v_i$ or $v_j$.

We leverage the pre-calculated behavior-aware semantic similarities $\beta_{i,j}$ between items for the first and second cases. Here $\beta_{i,j}$ is truncated from the top-$k$ value to be consistent with the semantic dependency hyperedges. We denote the values of the third case by $w_{i,j}$. Therefore, based on the above discussions, for the first case, we have: $\sum_{m=1}^{|\mathcal{E}^p|}\boldsymbol{\mathcal{M}}_{im}\boldsymbol{\mathcal{M}}_{jm} = \beta_{i,j} + w_{i,j}$; for the second case, we have $\sum_{m=1}^{|\mathcal{E}^p|}\boldsymbol{\mathcal{M}}_{im}\boldsymbol{\mathcal{M}}_{jm} = \beta_{i,j} + \beta_{j,i} + w_{i,j}$. Since the computation of the second-order complexity is time-consuming, and a number of such values are slight due to the top-$k$ truncation, we 
replace $w_{i,j}$ by a hyperparameter $w_0$ to obtain a close approximation $\boldsymbol{\mathcal{M}}^\prime$ for $\boldsymbol{\mathcal{M}}\boldsymbol{\mathcal{M}}^\trans$. Formally, $\boldsymbol{\mathcal{M}}^\prime \in \mathbb{R}^{J\times J}$ is generated as follows:
\begin{align}
    \boldsymbol{\mathcal{M}}^\prime = C \oplus A \oplus W
\end{align}
where $C_{i,j} = 1$ if $v_i=v_j$, otherwise $C_{i,j} = 0$, and $A_{i,j} = \beta_{i,j}$ if $v_i=v_j$, otherwise $A_{i,j} = \beta_{i,j} + \beta_{j,i}$. $W_{i,j} = w_0$. In our experiments, $w_0$ is searched among $[0.05, 0.1, 0.15, 0.2]$. Finally, we express the light version of hypergraph convolution function below:
\begin{align}
	\textbf{X}^{(l+1)} = \textbf{D}^{-1} \cdot \boldsymbol{\mathcal{M}}^\prime \cdot \textbf{X}^{(l)}
\end{align}


\begin{table}[t]
\caption{Comparison between two implementation of hypergraph message passing schemes, \ie the original embedding propagation based on hypergraph connection matrices, and the simplified message passing schema. The recommendation accuracy is measured by Recall@5 and model computational cost is measured by the running time of each epoch.}
\label{tab:lhgcn}
\resizebox{\linewidth}{!}{
\begin{tabular}{c|cc|cc|cc}
\hline
& \multicolumn{2}{c|}{Taoabo}                   & \multicolumn{2}{c|}{Retailrocket}             & \multicolumn{2}{c}{IJCAI}                    \\
& Recall@5 & Epoch Time& Recall@5 & Epoch Time & Recall@5 & Epoch Time\\
\hline
Origin     & 0.326                  & 86.43               & 0.959                  & 4.58                & 0.352                  & 133.28                \\
Simplified & 0.323                  &   38.05             & 0.956                  & 2.11                & 0.346                  & 65.2 \\      
\hline
\end{tabular}
}
\end{table}

We conduct empirical experiments to further evaluate the model performance of our simplified hypergraph convolution mechanism. We report the evaluation results in Table~\ref{tab:lhgcn}. The model training is performed on a single GTX3090 GPU for running time evaluation. We can observe that our simplified hypergraph-based message passing scheme only lead to slightly performance degradation and improve the model efficiency with lower computational cost. The potential reasons are: i) a large number of second-order similarity are slight, since scores are truncated from top-$k$ and ii) the convolution process inherently takes into account high-order connectivity. At the same time, simplifying the convolutional function brings a lot enhancement on inference speed, since it eliminates the original $O((|\mathcal{E}^p|+ |\mathcal{E}^q|)\times J^2)$ calculations and $\boldsymbol{\mathcal{M}}^\prime$ can be easily built by leveraging pre-calculated values and data pre-processing.


\begin{figure}[t]
\centering
\subfigure[Sensitivity to attention scales]{
\label{fig:HP:scale}
\includegraphics[width=0.48\linewidth]{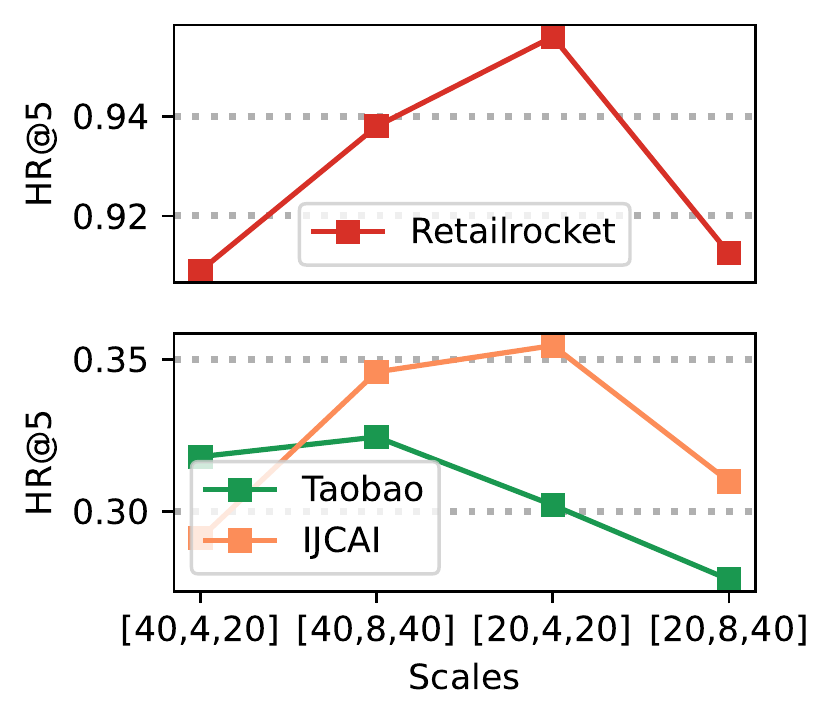}}
\subfigure[Sensitivity to sim-group length]{
\label{fig:HP:glen}
\includegraphics[width=0.45\linewidth]{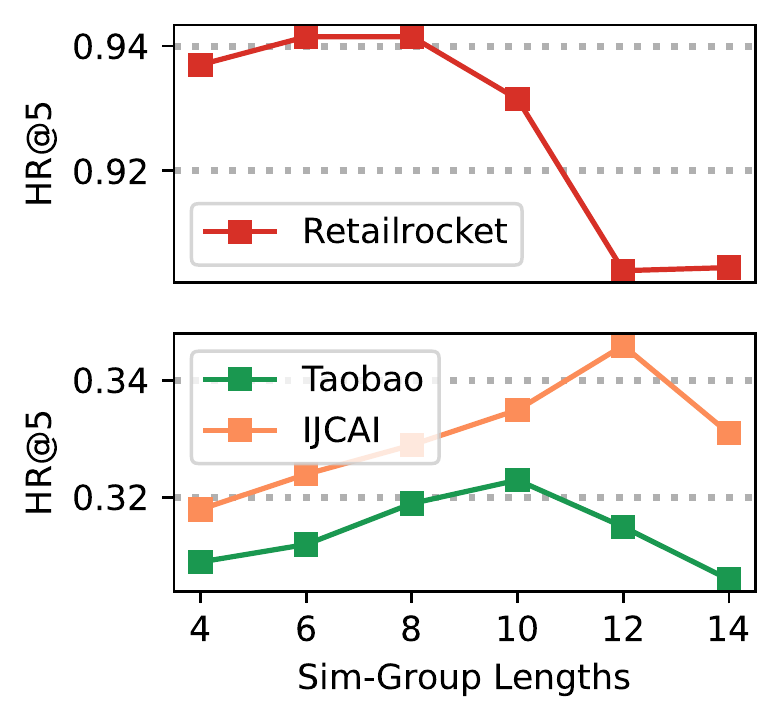}}
\vspace{-0.1in}
\caption{Hyperparameter study of \model\ framework.}
\label{fig:HP}
\end{figure}

\subsection{Hyperparameter Study (RQ4)}
\label{sec:hp}
We conduct experiments to analyze the influence of key hyperparameters in our \model\ framework and report results in Figure~\ref{fig:HP}.\\\vspace{-0.12in}

\noindent \textbf{Impact of Multi-Scale Settings}. We search multi-scale setting parameters $(C, p_1, p_2)$ among the range $\{[20,4,20],[20,8,40],[40,4,20]$ $[40,8,40]\}$. We present the observations as followed:
\begin{itemize}[leftmargin=*]
\item The best performance on Retailrocket and Taobao datasets can be achieved with $(p_1,p_2)=(4,20)$ and $(p_1,p_2)=(8,40)$ given the difference of average sequence length.\\\vspace{-0.1in}
\item For the low-rank projection scale parameter $C$, we can notice that projecting original self-attentive sequence embedding space into $\frac{J}{20}$ channels can bring the best performance on IJCAI and Retailrocket datasets. For Taobao dataset, we can observe that $\frac{J}{40}$ performs better than $\frac{J}{20}$, which indicates that dense user-item interaction data may need less low-rank projection channels for better sequential pattern encoding.
\end{itemize}


\noindent \textbf{Impact of Item-wise Semantic Dependency Set}. Our hypergraph item dependency encoder investigates the latent semantic correlations among different items. We search the top-$k$ semantic dependent items from \{4,6,8,10,12,14\} for global message passing through the item-wise semantic hyperedges. Observations are:
\begin{itemize}[leftmargin=*]
\item The best settings of $k$ is proportionally to the average sequence length of different datasets. It indicates that larger hypergraph propagation scope is better for modeling longer item sequences.\\\vspace{-0.1in}
\item Increasing the number of connected items through hyperedges may firstly boost the performance at the early stage, and then lead to performance degradation by involving noise during the hypergraph-based embedding propagation.
\end{itemize}

\end{document}